\documentclass[useAMS, usenatbib]{mn2e}
\usepackage{amsmath, amssymb, amsfonts}
\usepackage{bm}
\usepackage{graphicx}
\usepackage{placeins}
\usepackage{aas_macros}

\title{Tidal interactions of a Maclaurin spheroid. II: Resonant excitation of modes by a close, misaligned orbit}
\author[Harry J. Braviner and Gordon I. Ogilvie]
{Harry J. Braviner%
\thanks{
h.j.braviner@damtp.cam.ac.uk
}
and Gordon I. Ogilvie%
\thanks{
gio10@cam.ac.uk
}\\
Department of Applied Mathematics and Theoretical Physics, University of Cambridge, Centre for Mathematical Sciences,\\
Wilberforce Road, Cambridge, CB3 0WA}

\begin{document}

\maketitle

\begin{abstract}

We model a tidally forced star or giant planet as a Maclaurin spheroid, decomposing the motion into the normal modes found by \citet{Bryan1889}.
We first describe the general prescription for this decomposition and the computation of the tidal power.
Although this formalism is very general, forcing due to a companion on a misaligned, circular orbit is used to illustrate the theory.
The tidal power is plotted for a variety of orbital radii, misalignment angles, and spheroid rotation rates.
Our calculations are carried out including all modes of degree $l \le 4$, and the same degree of gravitational forcing.
Remarkably, we find that for close orbits ($a / R_* \approx 3$) and rotational deformations that are typical of giant planets ($e \approx 0.4$) the $l=4$ component of the gravitational potential may significantly enhance the dissipation through resonance with surface gravity modes.
There are also a large number of resonances with inertial modes, with the tidal power being locally enhanced by up to three orders of magnitude.
For very close orbits ($a / R_* \approx 3$), the contribution to the power from the $l=4$ modes is roughly the same magnitude as that due to the $l=3$ modes.

\end{abstract}


\section{Introduction}
\label{sec:introduction}

The recent discovery of a myriad of exoplanets by Kepler, WASP, and other projects has uncovered a large number having very short period orbits around their host stars.
At the time of writing the Exoplanet Orbit Database (\citet{exoplanets}) lists over 200 planets with orbital periods of less than 5 days, 74 of which have masses greater than Jupiter.
In these systems the tidal interactions between the planet and the host star are expected to be significant, and potentially responsible for the circularisation and spin-orbit alignment of many systems.
The observational evidence for such tidal evolution has been examined in \citet{JGB2008}, \citet{Hansen2010}, \citet{Husnoo2012}, and \citet{Albrecht2012}.
Circularisation provides strong evidence for the action of tides within the planet, but is less sensitive to the tidal dissipation in the host star.
The dichotomy between stars cooler than $6250\,\mathrm{K}$, which host mainly aligned planets, and stars hotter than $6250\,\mathrm{K}$, amongst which the distribution of hot Jupiter orbits is nearly isotropic, has been interpreted as evidence of tidal interaction.

Dissipation of the stellar tide may also cause hot Jupiters to migrate radially inwards, decreasing their orbital period.
Over many orbits, a small change in the period could cause a measurable change in the time at which the transit occurs.
\cite{BCC2014} discuss this migration, and argue that a stellar tidal quality factor, $Q^{\prime}_*$, of $10^6$ would produce measurable transit timing shifts in some planets after 15 years ($17\, \mathrm{s}$ for WTS-2, $801\,\mathrm{s}$ for WASP-18b; the current timing accuracy is given as $5\,\mathrm{s}$).
\citeauthor{BCC2014} suggest that inferring constraints on $Q^{\prime}_*$ from the known population of hot Jupiters is subject to a large degree of uncertainty regarding the sensitivity of transit surveys and whether the planets have undergone disc-driven migration or Lidov-Kozai scattering.
They also show, however, that this population can provide constraints on $Q^{\prime}_*$ down to $10^7$ after 10 years of transit timing observations, with independent constraints for each star.
The independence of these constraints is important;
stars more massive than $1.25 M_{\odot}$ lack the surface convective layers that are thought to contribute strongly to the tidal dissipation in solar-type stars, and are expected to have a higher $Q^{\prime}_*$ as a result.
\cite{MPZ2014} measured a period variation of $-0.15 \pm 0.06 \,\mathrm{sec} / \mathrm{year}$ for WASP-43b.
This is smaller than the previous measurement of \cite{BHM2014}, and \citeauthor{MPZ2014} suggest that additional observations over a longer time span are needed to truly confirm this orbital decay.

Aside from direct measurement of orbital decay, the presence of hot Jupiters with semi-major axes less than twice their Roche limit, $a_{\mathrm{R}}$, hints at the action of stellar tides.
One formation mechanism for hot Jupiters involves the orbital eccentricity of the planet being driven to high values by the Lidov-Kozai mechanism.
The periapsis of the planet must not become less than $a_{\mathrm{R}}$ during this process, or the planet will be destroyed.
The circularisation of the orbit is assumed to be dominated by the planetary tide, which will not to change the orbital angular momentum.
This leads to a circular orbit of radius $2 a_{\mathrm{R}}$.
\cite{VR2014} used a tidal model including both a weak-friction equilibrium tide model of \cite{VR2014b} and the inertial wave model of \cite{Lai2012} and found that tidal migration was able to explain the presence of hot Jupiter so close to their host-stars.


In this work we investigate the tidally forced flows in a giant planet or low mass star, modelled as linear flows in a Maclaurin spheroid.
This is a body of homogeneous, incompressible fluid in a state of solid body rotation.
Such a model supports surface gravity and inertial waves, but not internal gravity waves.
It is therefore a better approximation to convective, rather than radiative, regions.
We shall include forcing by gravitational harmonics up to degree $l=4$, since this increases the spectrum of inertial modes which may be resonantly excited.
These may be particularly significant for short-period misaligned planets and satellites.
An additional advantage of this model comes from its inclusion of an equatorial bulge due to rotation, and the fact that its normal modes were found analytically by \citet{Bryan1889}.
These features make the Maclaurin spheroid an attractive (and computationally cheap) option for studying the effects of rapid rotation on the tidal interaction of a giant planet or low mass star.
This model could also be used to study some aspects of tides in terrestrial bodies.

In \cite{BO2014} (hereafter `paper I') we reviewed the work of \citet{Bryan1889} and \citet{LI1999} and showed that all the normal modes of a Maclaurin spheroid (including the special cases of frequency $0$, $2 \Omega$ and $-2 \Omega$) could be labelled by their degree ($l$), order ($m$) and frequency ($\kappa \Omega$).
We numerically calculated the frequencies and decay rates of these modes up to $l=4$, and included plots showing the spatial form of these flows.

In this paper we shall consider the problem of forced oscillations, which we illustrate with the case of a planet in a circular orbit that may be misaligned with respect to the rotation axis of the star.
We choose to study circular orbits partly for their simplicity, but also because the circularisation of hot Jupiters is expected to occur much more rapidly than alignment or migration, leaving many planets in this configuration.
Since this is a linear theory, we begin by showing how to decompose the response to a single Fourier mode of forcing in section \ref{sec:forced_problem} and appendix \ref{sec:decomposition}.
We consider the potential due to the misaligned orbit in section \ref{sec:oblique_orbit_potential}, and show how this is related to the tidal power in section \ref{sec:oblique_orbit_power} and appendix \ref{sec:power_torque_calculation}.
Also in section \ref{sec:oblique_orbit_power}, we compute the tidal power due to orbits of various misalignments, at orbital radii of three and six stellar radii, as the eccentricity of the spheroid is varied.
Such a plot may be thought of as describing how the tidal dissipation varies as a young star spins down due to magnetic braking.
We also plot the power versus orbital radius for spheroids of fixed eccentricities $e=0.2$ and $0.4$.
These plots show the variations in dissipation rate as a planet migrates radially due to the action of the tide.
We study the power, rather than the torque, as a simple measure of the interaction.
The torque is a vector, and is not positive-definite.
While the torque is required to understand the orbital evolution, we postpone computing this to later work.
In section \ref{sec:oblate_Love_numbers} we show that, subject to a small approximation, the response of the spheroid may be described by a set of potential Love numbers.
These are not the usual Love numbers of spherical harmonics, but of the oblate spheroidal harmonics introduced in paper 1.

\section{The Forced Response of The Maclaurin Spheroid}
\label{sec:forced_problem}

\begin{figure}
\centering
\input{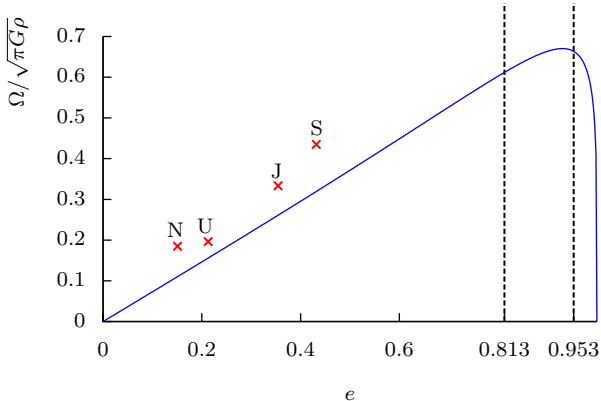}
\caption{
Non-dimensionalised angular velocity ($\Omega$) for a Maclaurin spheroid against eccentricity ($e$).
For $e \gtrsim 0.8127$ the spheroid has a secular instability and for $e \gtrsim 0.9529$ it has a dynamical instability.
We have marked the Solar System gas and ice giants, according to their mean densities.
}
\label{fig:maclaurin_spheroid_angular}
\end{figure}

A Maclaurin spheroid is a mass of homogeneous, incompressible fluid, of density $\rho$, in solid body rotation occupying the oblate spheroidal volume $V$ given by
\begin{equation}\label{spheroid_cylindricals} \frac{x^2 + y^2}{R_e^2} + \frac{z^2}{R_p^2}
= \frac{\varpi^2}{R_e^2} + \frac{z^2}{R_p^2}
\le 1 \,,
\end{equation}
where the equatorial and polar radii obey $R_e > R_p$, and $\varpi$ is cylindrical polar radius measured from the $z$ axis.
Following the convention of paper I, we specify the shape of the spheroid by its eccentricity, $e = \left(1 - (R_p/R_e)^2\right)^{1/2}$, and its overall scale by the mean radius, $R = (R_e^2 R_p)^{1/3}$.
We will sometimes use the focal radius, $c = e R_e$, and the parameter $\zeta_0 = \sqrt{1-e^2}/e$, for brevity.
Maclaurin spheroids exist for rotation rates $\Omega \lesssim 0.6703 \sqrt{\pi G \rho}$, with two eccentricities corresponding to each rotation rate.
However, the entirety of the high eccentricity branch of solutions is unstable for a spheroid composed of viscous fluid.
We plot the relation between the angular velocity and eccentricity in figure \ref{fig:maclaurin_spheroid_angular}.
We note that this relation will differ from that of a real, inhomogeneous body.
We have illustrated this by plotting the Solar System gas and ice giants, taking $\rho$ to be their mean density and defining the eccentricity of their figure via $R_p / R_e$.
More details about the Maclaurin spheroid can be found in paper I.

\citet{Bryan1889} found the normal modes of such a spheroid of inviscid fluid, solving the linearised Euler equation for the velocity perturbation, $\bm{u}$, in the frame rotating with the spheroid,
\begin{align}\label{inviscidNS}
\partial_t \bm{u} + 2 \bm{\Omega} \times \bm{u} = - \bm{\nabla} W \;\;
\mathrm{and} \;\; \bm{\nabla}\cdot \bm{u} = 0 \;\; \mathrm{inside} \; V \,.
\end{align}
The perturbations to the pressure and gravitational potential have been collected into the hydrodynamic potential, $W = p^{\prime} / \rho + \Phi^{\prime}$.
This is subject to the boundary condition that the pressure vanish at the displaced surface,
\begin{align} \label{pbc}
p^{\prime} + \bm{\xi} \cdot \bm{\nabla} p = 0 \; \mathrm{on} \; \partial V \,,
\end{align}
where $\bm{\xi}$ is the displacement, obeying $\bm{u} = \partial_t \bm{\xi}$.
The perturbation to the gravitational potential, $\Phi^{\prime}$, must obey the Poisson equation sourced by the Eulerian density perturbation.
In the linear theory this reduces to Laplace's equation away from the surface,
\begin{align} \label{Poisson_eqn}
\nabla^2 \Phi^{\prime} = 0 \; \mathrm{away}\;\mathrm{from}\; \partial V \,,
\end{align}
and a boundary condition due to a surface mass density source,
\begin{align}\label{Phibc}
\left[\bm{n} \cdot \bm{\nabla} \Phi^{\prime}\right]^{\partial V^+}_{\partial V^-} &= 4 \pi G \rho \bm{n} \cdot \bm{\xi} \;\; \mathrm{across} \; \partial V \,.
\end{align}

In our previous paper we reviewed the derivation of the normal modes, largely following the route of \citet{LI1999} and being careful to cover some special frequency cases that had not been considered previously.
Our convention for the frequency is that all perturbations have time dependence $e^{-i\kappa \Omega t}$.
We extended this work by seeking solutions to this problem for a weakly viscous fluid, replacing \eqref{inviscidNS} by
\begin{align}\label{viscousNS}
\partial_t \bm{u} + 2 \bm{\Omega} \times \bm{u} = - \bm{\nabla} W + \nu \nabla^2 \bm{u} \,.
\end{align}
In the inviscid problem the spheroid is known to be dynamically stable (see \citet{Chandrasekhar1987}) for $e \lesssim 0.9529$ and therefore the values of $\kappa$ found in this range are purely real, representing oscillatory solutions.
We treated the viscosity of \eqref{viscousNS} as a perturbation to the original problem and found that this introduced an $\mathcal{O}(\nu)$ imaginary part to $\kappa$.
For $e \lesssim 0.8127$ the imaginary part of $\kappa$ is always found to be negative, corresponding to a decay rate for the mode.
For $e$ greater than this threshold some mode frequencies have a positive imaginary part, corresponding to a viscous growth rate, in accordance with the spheroid being secularly unstable for these eccentricities.

In this paper we first wish to solve the problem of a viscous fluid subject to a harmonic potential force, presumed to originate from the gravitational field of an orbiting companion%
\footnote{
The tidal forcing will in fact consist of several frequencies, as we shall see in section \ref{sec:oblique_orbit_potential}.
However, since we are only considering the linear theory, these may be treated independently.
}.
We replace \eqref{inviscidNS} by
\begin{align}\label{viscousforcedNS}
\partial_t \bm{u} + 2 \bm{\Omega} \times \bm{u} = - \bm{\nabla} W + \nu \nabla^2 \bm{u} + \bm{f} e^{-i \kappa \Omega t} \,,
\end{align}
where $\bm{f} e^{-i \kappa \Omega t} = - \bm{\nabla} \phi$ is the forcing due to the gravity of the companion.
The boundary conditions \eqref{pbc} and \eqref{Phibc} are unchanged.

Any solution of \eqref{viscousNS} will also satisfy this problem, but will decay in amplitude on the viscous timescale.
We restrict ourselves to seeking solutions with the same harmonic time dependence as the forcing potential $\phi$, assuming that any transients decay on a timescale that is short enough that they do not significantly affect the tidal evolution.
Denoting the spatially dependent factor of Bryan's inviscid free modes by $\bm{\xi}_{\alpha}$, we seek a decomposition of the forced solution as
\begin{align} \label{LambdaExpDef}
\left( \begin{array}{c} \bm{\xi} \\ \bm{u} \end{array} \right)
= \sum_{\alpha} \Lambda_{\alpha} \left( \begin{array}{c} \bm{\xi}_{\alpha} \\ \bm{u}_{\alpha} \end{array} \right)
e^{-i \kappa \Omega t} \,.
\end{align}
The index $\alpha$ labels distinct modes.
It may be viewed as an shorthand for the degree ($l$), order ($m$) and frequency ($\kappa^{\prime}$) of a free mode.
Since we shall truncate our expansion at modes of degree $l=4$, we are not capable of resolving the thin free-slip boundary layer that will exist near the surface of the star.
The dissipation due to a free-slip boundary layer of thickness $\delta$ is typically $\mathcal{O}\left( \delta / R \right)$ relative to the bulk flow of length scale $R$, therefore the contribution of this layer to the dissipation may be neglected.

In appendix \ref{sec:decomposition} we show that the expansion coefficients must obey
\begin{align} \label{LambdaForcingRelationMainText}
\Omega \left(\kappa - \kappa_{\beta} - \delta \kappa_{\beta} \right) \epsilon_{\beta} \Lambda_{\beta} 
+ & 2 i \nu \sum_{\alpha \ne \beta} \left( \int_V e_{\beta i j}^* e_{\alpha i j} \mathrm{d}V \right) \Lambda_{\alpha} \nonumber \\
& = i \int_V \bm{u}_{\beta}^* \cdot \bm{f} \mathrm{d}V \,.
\end{align}
Here
\begin{align} \label{epsilonDef}
\epsilon_{\alpha} = \int_V \bm{u}_{\alpha}^* \cdot \bm{u}_{\alpha} \mathrm{d} V + \int_V \bm{\xi}_{\alpha}^* \cdot \bm{\nabla} W_{\alpha} \mathrm{d} V
\end{align}
is proportional to the energy of the inviscid mode, $\kappa_{\alpha} \Omega$ is the inviscid frequency of a free mode, and $\delta \kappa_{\alpha}$ is the viscous perturbation to $\kappa_{\alpha}$ given by
\begin{align}
\label{deltakappaDef}
\delta \kappa_{\alpha} = -2 i \nu \int_V e_{\alpha i j}^* e_{\alpha i j} \mathrm{d}V \bigg/ \Omega \epsilon_{\alpha} \,.
\end{align}
$e_{\alpha i j} = \frac{1}{2} \left( \partial_i u_{\alpha j} + \partial_j u_{\alpha i} \right)$ is the rate-of-strain tensor of the inviscid mode.

In principle \eqref{LambdaForcingRelationMainText} involves all of the free modes of the spheroid.
In practice we may impose a cut-off at some finite degree by noting two properties of this expression.
Firstly, the gravitational forcing potential may be expanded in harmonics of the oblate spheroidal coordinates introduced in paper I, as shown in \eqref{phiOblateExpansion}.
The integral $\int_V \bm{u}_{\beta}^* \cdot \bm{f} \mathrm{d} V$ is proportional to some expansion coefficient $\phi_{(o)}{}_l^m$ with the same degree, $l$, and order, $m$, as the free mode labelled by $\beta$.
Such coefficients scale with orbital radius, $r_{\mathrm{orbit}}$, and the mean radius of the spheroid, $R$, like $\phi_{(o)}{}_l^m = \mathcal{O} \left( \left( R / r_{\mathrm{orbit}} \right)^l \right)$.
Secondly, the `mixing' integral between different modes, $\int_V e^*_{\alpha i j} e_{\beta i j} \mathrm{d} V$, depends upon the magnitude of the rate-of-strain tensor.
As we increase the degree of the mode we expect the components of this tensor to possess an increasing number of sign changes with radius (compare illustrations of the $l=4$ and $l=3$ flows given in paper I).
If $\alpha$ labels a mode of low degree, and $\beta$ a mode of high degree, there should be a high degree of cancellation between these and the integral is expected to be small.
Therefore we believe that high order modes cannot be significantly excited;
the corresponding mode of the gravitational potential is not strong enough to excite them directly, nor can they be significantly excited by viscous interaction with a strongly forced mode of lower order.

When we wish to numerically evaluate quantities in this paper we shall restrict our attention to the modes of degree $l \le 4$.
We shall not yet discard the mode mixing term $2 i \nu \int_V e^*_{\alpha i j} e_{\beta i j} \mathrm{d} V$ between the retained modes, though in section \ref{sec:oblate_Love_numbers} we will argue that only the mixing between the $l=2$ surface gravity modes is significant.

To make quantitative statements about the evolution of the orbit of a satellite of such a spheroid we must calculate the time-averaged power and the time-averaged $z$ component of the torque associated with the tidal interaction.
These may be defined by
\begin{align}
\left< P \right> & = \rho \int_V \left< \mathrm{Re}\left(\bm{u}\right) \cdot \mathrm{Re}\left(- \bm{\nabla} \phi\right) \right> \mathrm{d} V \\
\mathrm{and} \,\, \left< T_z \right> & = \rho \int_{\partial V} \left< \mathrm{Re} \left( \bm{\xi} \cdot \bm{n} \right) \mathrm{Re} \left( - \partial_{\varphi} \phi \right) \right> \mathrm{d} S \,.
\end{align}
In the case of a single mode of order $m$ forced at frequency $\kappa \Omega$, the latter of these is equivalent to $\left< T_z \right> = m \left< P \right> / \kappa \Omega$.

However, we shall first consider the potential produced by this misaligned orbit, in section \ref{sec:oblique_orbit_potential}, before performing this power calculation in section \ref{sec:oblique_orbit_power}.

\section{The Potential of a Point-Mass on an Oblique, Circular Orbit}
\label{sec:oblique_orbit_potential}

Let the companion be of mass $M_2$ and in a circular orbit of radius $r_{\mathrm{orbit}}$ with inclination $\theta_{i}$ to the equatorial plane of the Maclaurin spheroid.
We denote the angular frequency of the orbit by%
\footnote{%
We have used the notation $r_{\mathrm{orbit}}$ and $\omega_{\mathrm{orbit}}$ in preference to the more usual $a_{\mathrm{orbit}}$ and $n_{\mathrm{orbit}}$ to avoid misleading the reader into thinking that these results apply to an eccentric orbit.
}
$\omega_{\mathrm{orbit}}$.
We first expand the gravitational field of $M_2$ in spherical harmonics with respect to a spherical polar coordinate system $(r, \theta^{\prime \prime}, \varphi^{\prime \prime})$, the $\theta^{\prime \prime}=0$ axis of which is aligned with the angular momentum of the orbit.
This coordinate system is not rotating.
We may write
\begin{align}
\phi = - \frac{G M_2}{\left| \bm{r}_2 - \bm{r}\right|}
= - \frac{G M_2}{r_{\mathrm{orbit}}} \sum_{l=0}^{\infty} \left( \frac{r}{r_{\mathrm{orbit}}} \right)^l P_l \left( \cos \gamma \right) \,,
\end{align}
where $\gamma$ is the angle between the location of the planet, $(r_{\mathrm{orbit}}, \pi/2, \omega_{\mathrm{orbit}} t)$, and the point $(r,\theta^{\prime \prime}, \varphi^{\prime \prime})$ at which we wish to evaluate the potential.
Using the addition theorem for spherical harmonics (see \citet{Jackson1975}) and the fact that the associated Legendre polynomials $P_l^m(0)$ are non-zero only for even $l+m$, we may write
\begin{align} \label{phiInertialFrame}
\phi
= & - \frac{G M_2}{r_{\mathrm{orbit}}} \sum_{l=0}^{\infty} \Bigg[ \left( \frac{r}{r_{\mathrm{orbit}}} \right)^l \frac{4 \pi}{2l + 1} \nonumber \\
& \hspace{1.7cm} \times \sum_{m=-l}^l Y_l^{m *} (\pi/2, \omega_{\mathrm{orbit}} t ) Y_l^m ( \theta^{\prime \prime}, \varphi^{\prime \prime} ) \Bigg] \nonumber \\
= & - \frac{G M_2}{r_{\mathrm{orbit}}} \Bigg[ \sum_{l=0}^{\infty} \left( \frac{r}{r_{\mathrm{orbit}}} \right)^l \sqrt{\frac{4 \pi}{2l+1}} \nonumber \\
\times & \sum_{\substack{m = -l \\ m+l\; \mathrm{even}}}^{l} \left(-1\right)^m
P_l^m\left( 0 \right)
\sqrt{\frac{\left(l-m\right)!}{\left(l+m\right)!}} Y_l^m (\theta^{\prime \prime}, \varphi^{\prime \prime}) e^{-i m \omega_{\mathrm{orbit}} t} \Bigg] \nonumber \\
= & \sum_{l=0}^{\infty} \sum_{m = -l}^l \phi_{(s)}^{\prime \prime}{}_l^m \left(\frac{r}{R}\right)^l Y_l^m( \theta^{\prime \prime}, \varphi^{\prime \prime} ) \, .
\end{align}
In the final line we have packaged this sum up into the coefficients $\left\{ \phi_{(s)}^{\prime \prime}{}_l^m \right\}$.
The mean radius, $R$, is a convenient length by which to non-dimensionalise, since it does not change with either the evolution of the orbit or the angular momentum of the spheroid.

We now wish to convert to a non-rotating coordinate system $(r, \theta, \varphi^{\prime})$, the $\theta=0$ axis of which is aligned with the angular momentum of the spheroid.
Without loss of generality we assume that the ascending node of the orbit lies in the $\widehat{\bm{y}}^{\prime}$ direction.
We may do so because of the azimuthal symmetry of the spheroid and by assuming that transients decay on a timescale much shorter than the nodal precession of the orbit.
The coefficients in these coordinates are
\begin{align} \label{littledtrans}
\phi_{(s)}^{\prime}{}_l^m = \sum_{m^{\prime}= - l}^{l} d^{(l)}_{m m^{\prime}}(\theta_i) \phi_{(s)}^{\prime \prime}{}_l^{m^{\prime}} \,,
\end{align}
where $d^{(l)}_{m m^{\prime}}$ are the elements of Wigner's small $d$-matrices.
They are given by
\begin{align}
d^{(l)}_{m m^{\prime}} \left( \theta_i \right) = & \sqrt{\frac{(l-m^{\prime})!(l+m)!}{(l+m^{\prime})!(l-m)!}} \nonumber \\
& \times \frac{\left( \cos \theta_i / 2 \right)^{2l + m^{\prime} -m} \left( - \sin \theta_i / 2 \right)^{m - m^{\prime}}}{(m - m^{\prime})!} \nonumber \\
& \times {_2}F_1\left(m-l, -m^{\prime}-l; m-m^{\prime}+1; - \tan^2 \frac{\theta_i}{2} \right) \,,
\end{align}
for $m \ge m^{\prime}$, where ${_2}F_1$ is the hypergeometric function.
The case of $m \le m^{\prime}$ may be obtained via the relation
$d^{(l)}_{m^{\prime} m} \left(\theta_i\right) = \left(d^{(l)}_{m m^{\prime}} \left(-\theta_i\right)\right)^* $.
Other conventions for the $d$-matrices exist%
\footnote{In particular, be aware that in the computer algebra system \emph{Mathematica} the function $\texttt{WignerD}\left[\left\{ l, m, m^{\prime} \right\}, 0, \theta_i, 0\right]$ is equal to our $d^{(l)}_{m m^{\prime}} \left(-\theta_i\right)$. };
we use that of \citet{Morrison1987}, in which a clear explanation of these matrices and their relation to spherical harmonics may be found.

Finally we transform to the coordinate system $(r, \theta, \varphi)$ aligned with and rotating with the angular velocity of the Maclaurin spheroid.
Since $\varphi = \varphi^{\prime} - \Omega t$, we have
\begin{align} \label{phiprimetophitrans}
\phi_{(s)}{}_l^m = \sum_{m^{\prime} = -l}^l d^{(l)}_{m m^{\prime}}(\theta_i) e^{i m \Omega t} \phi_{(s)}^{\prime \prime}{}_l^{m^{\prime}} \,.
\end{align}

To find the response of the spheroid we must express the potential in terms of the oblate spheroidal harmonic coefficients, $\left\{ \phi_{(o)}{}_l^m \right\}$, defined by
\begin{align}
\phi = \sum_{l=0}^{\infty} \sum_{m=-l}^l \phi_{(o)}{}_l^m \frac{P_l^m(\mu) P_l^m(i\zeta)}{P_l^m(i\zeta_0)} e^{i m \varphi} \,.
\end{align}
To do so requires us to rewrite $P_l^m(i \zeta) P_l^m(\mu) e^{i m \varphi}$ as a sum over terms of the form $\left(r / c\right)^{l^{\prime}} P_{l^{\prime}}^{m}(\cos \theta) e^{i m \varphi}$.
Here $\zeta$ and $\mu$ denote the oblate spheroidal coordinates, defined by
\begin{align} \label{oblateSpherCoordDef}
\varpi^2 =  c^2 \left(1 + \zeta^2\right) \left(1 - \mu^2\right), \;\; z = c \zeta \mu \,.
\end{align}
These are described in detail in appendix A of paper I.
Clearly such an expansion is possible - the oblate spheroidal harmonics solve Laplace's equation and are regular at the origin, and hence must possess an expansion in terms of interior spherical harmonics, and the azimuthal dependence must match.
Defining $\mathrm{H}^{(m)}_{l l^{\prime}}$ by
\begin{align}
P_l^m(i \zeta) P_l^m(\mu) = \sum_{l^{\prime} = 0}^{\infty} \mathrm{H}^{(m)}_{l l^{\prime}} \left(r / c\right)^{l^{\prime}} P_{l^{\prime}}^m \left(\cos \theta\right) \,,
\end{align}
we may write the coefficients of the expansion with respect to oblate harmonics, using the matrix inverse of $\bm{\mathrm{H}}^{(m)}$, as
\begin{align} \label{sphertooblatetrans}
\phi_{(o)}{}_l^m = & (-1)^m P_l^m(i \zeta_0) \nonumber \\
& \times \sum_{l^{\prime}=0}^{\infty} \sqrt{\frac{2 l^{\prime} + 1}{4\pi} \frac{(l^{\prime}-m)!}{(l^{\prime} + m)!}}\left(\frac{c}{R}\right)^{l^{\prime}}
\mathrm{H}^{(m) -1}_{l^{\prime} l}
\phi_{(s)}{}_{l^{\prime}}^m \,.
\end{align}
We tabulate the $\bm{\mathrm{H}}^{(m)}$ matrices in appendix \ref{sec:oblate_spherical_harmonic_conversions}, from which we see that both of the preceding sums receive contributions only from $l^{\prime} \ge l$.

Let us now consider to what extent the spectrum of forcing frequencies is enhanced by these transformations.
In combination with the mode frequencies found in paper I, this will allow us to plot the orbital frequencies at which resonances may occur.
Symmetry about the equatorial plane shows that an aligned orbit is only able to excite two of the five $l=3$ inertial modes, and four of the eight $l=4$ inertial modes (this may be seen by considering the parity of $l+m$ for these modes).
For a misaligned orbit the transformation \eqref{littledtrans} mixes up the coefficients $\phi_{(s)}{}_l^m$ of the same degree, $l$, but differing order, $m$.
For a generic misalignment, all orders within a given degree may be excited with comparable magnitudes.
This should increase the opportunities available for resonant tidal forcing.

Further, the spectrum of forcing frequencies in the frame rotating with the spheroid is much greater than that of the inertial frame.
We see from \eqref{phiInertialFrame} that $\phi_{(s)}^{\prime \prime}{}_l^m$, the coefficients in the inertial frame aligned with the orbit, have harmonic time dependence with frequency $m \omega_{\mathrm{orbit}}$.
From \eqref{littledtrans} we see that, for a generic spin-orbit misalignment, the coefficients in the inertial frame aligned with the rotation axis of the spheroid, $\phi_{(s)}^{\prime}{}_l^m$, will contain Fourier modes with frequencies $\left\{ \left. m^{\prime} \omega_{\mathrm{orbit}} \; \right| \; \mathrm{for} \; m^{\prime} = -l, -l+2, \dots, l-2, l\right\}$.
The transformation \eqref{phiprimetophitrans} to the frame rotating with the spheroid introduces a Doppler shift, giving $\phi_{(s)}{}_l^m$ components of comparable magnitudes with frequencies in $\left\{ \left. m^{\prime} \omega_{\mathrm{orbit}} - m \Omega \; \right| \; \mathrm{for} \; m^{\prime} = -l, -l+2, \dots, l-2, l\right\}$.
Finally, the transformation from spherical to oblate spheroidal harmonics introduces into $\phi_{(o)}{}_l^m$ contributions from $\phi_{(s)}{}_l^m, \, \phi_{(s)}{}_{l+2}^m, \, \phi_{(s)}{}_{l+4}^m, \, \dots$. 
Therefore $\phi_{(o)}{}_l^m$ will contain components of frequencies $\left\{ \left. m^{\prime} \omega_{\mathrm{orbit}} - m \Omega \; \right| \; m^{\prime} \; \mathrm{such} \; \mathrm{that} \; l+m^{\prime} \; \mathrm{is} \; \mathrm{even.} \right\}$.
However, the new components that are introduced by this transformation are smaller by at least one power of $\left( c / r_{\mathrm{orbit}} \right)^{2}$
and so will only be significant for very close orbits around highly eccentric Maclaurin spheroids.

\begin{figure}
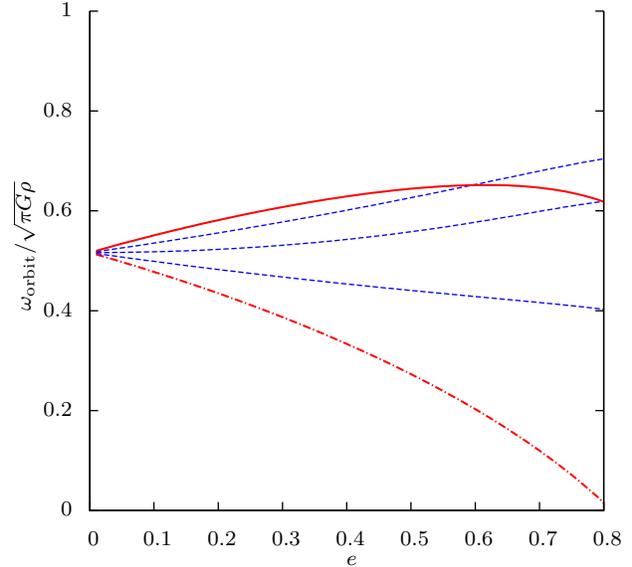

\include{figures/resonant_periods/l2resonantPeriods/l2resonantOrbitalFreqs}
\caption{
Orbital frequencies ($\omega_{\mathrm{orbit}}$), non-dimensionalised by the dynamical frequency of the Maclaurin spheroid ($\sqrt{\pi G \rho}$), at which the $l=2$ component of the gravitational potential of the point-mass companion may resonantly excite a mode of the spheroid.
The heavy red lines denote the resonances of an equatorial orbit, with the solid line applying to a prograde equatorial orbit, and the dash-dotted line applying to a retrograde equatorial orbit.
The lighter blue dashed lines denote the additional resonant frequencies that exist for a generic spin-orbit misalignment (owing to the Wigner $d$-matrix transformation \eqref{littledtrans}).
Such a misaligned orbit will typically also excite resonances at the frequencies shown by the heavy red lines, though the reader should note that the amplitude of each gravitational harmonic will depend on the misalignment;
some misaligned orbits will still fail to excite certain resonances due to not forcing these modes strongly.
}
\label{fig:l2resonantOrbitalFreqs}
\end{figure}

\begin{figure}
\include{figures/resonant_periods/l3resonantPeriods/l3resonantOrbitalFreqs}
\caption{
Orbital frequencies ($\omega_{\mathrm{orbit}}$), non-dimensionalised by the dynamical frequency of the Maclaurin spheroid ($\sqrt{\pi G \rho}$), at which the $l=3$ component of the gravitational potential of the point-mass companion may resonantly excite a mode of the spheroid.
As in figure \ref{fig:l2resonantOrbitalFreqs}, heavy red lines apply to an equatorial orbit (solid for prograde, dash-dotted for retrograde) and lighter blue dashed lines are the additional resonances encountered by a generically misaligned orbit.
}
\label{fig:l3resonantOrbitalFreqs}
\end{figure}

\begin{figure}
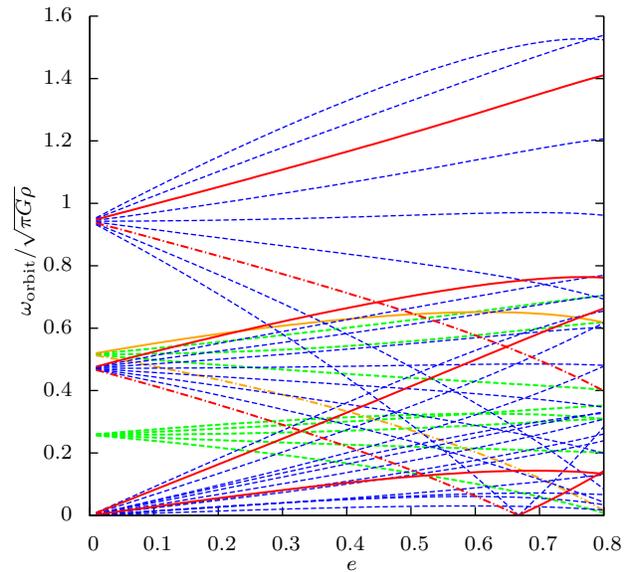

\include{figures/resonant_periods/l4resonantPeriods/l4resonantOrbitalFreqs}
\caption{
Orbital frequencies ($\omega_{\mathrm{orbit}}$), non-dimensionalised by the dynamical frequency of the Maclaurin spheroid ($\sqrt{\pi G \rho}$), at which the $l=4$ component of the gravitational potential of the point-mass companion may resonantly excite a mode of the spheroid.
As in figure \ref{fig:l2resonantOrbitalFreqs}, heavy red lines apply to an equatorial orbit (solid for prograde, dash-dotted for retrograde) and lighter blue dashed lines are the additional resonances encountered by a generically misaligned orbit.
In addition, the fact that a given spherical harmonic corresponds to a sum over multiple oblate harmonics allows this component of the gravitational potential to resonate with modes of oblate spheroidal degree $l=2$, as determined by \eqref{sphertooblatetrans}.
Orange lines show the frequencies of an equatorial orbit that may resonate with such modes, and green dashed lines show the corresponding frequencies for a general spin-orbit misalignment.
}
\label{fig:l4resonantOrbitalFreqs}
\end{figure}

We summarise in figures \ref{fig:l2resonantOrbitalFreqs}, \ref{fig:l3resonantOrbitalFreqs} and \ref{fig:l4resonantOrbitalFreqs} the orbital frequencies at which this point-mass companion may resonantly excite a mode of the spheroid with an amplitude of order $\left( R / r_{\mathrm{orbit}} \right)^l $, for $l=2$, $3$ and $4$ respectively.
These figures distinguish between resonant frequencies for a (possibly retrograde) equatorial orbit and an orbit with a general misalignment.
The equatorial orbit may only excite modes of the spheroid for which $l+m$ is even.
The misaligned orbit may excite all modes of the same degree, $l$.
In addition, the misalignment of the orbit mixes the frequencies between gravitational harmonics of different orders.
For example, the $\phi_{(s)}^{\prime}{}_2^2$ component for an aligned orbit will have time dependence $e^{-2i\omega_{\mathrm{orbit}}t}$, whereas the corresponding component for a misaligned orbit will have components with time dependence $e^{- 2i\omega_{\mathrm{orbit}}t}$ and $e^{+ 2i\omega_{\mathrm{orbit}}t}$ (in addition to a constant secular component that we do not consider to resonate).

The $\phi_{(s)}{}_4^m$ component of the spherical harmonic expansion of the gravitational potential also contributes to the $\phi_{(o)}{}_2^m$ component of the oblate harmonic expansion of this potential, as determined by \eqref{sphertooblatetrans} and appendix \ref{sec:oblate_spherical_harmonic_conversions}.
The resonances thus excited are distinguished from the other frequencies plotted in figure \ref{fig:l4resonantOrbitalFreqs}.
Some of these frequencies coincide with those plotted in figure \ref{fig:l2resonantOrbitalFreqs}, and in such cases the contribution from the $l=2$ spherical gravitational harmonic, with its $\left( R / r_{\mathrm{orbit}} \right)^2$ scaling, will dominate.
Note also that the amplitude of this forcing will scale with the focal radius of the spheroid, $c$, as $\left( c / R \right)^2$, and will therefore be of little importance in the spherical limit.
However, for a misaligned orbit, the $\phi^{\prime \prime}_{(s)}{}_4^m$ coefficients with $\left| m \right| > 2$, with their higher-frequency $e^{-im \omega_{\mathrm{orbit}}t}$ time-dependence, do allow lower orbital frequencies to resonate with the $l=2$ surface gravity modes.
Such frequencies are visible as the set of curves emerging from $\omega_{\mathrm{orbit}} \approx 0.26 \sqrt{\pi G \rho}$ in figure \ref{fig:l4resonantOrbitalFreqs}.
We shall see in the following section an example of such a resonance significantly enhancing the dissipation for a retrograde orbit.
In principle, the $l=5$ components of the spherical harmonic expansion of the gravitational potential should resonantly excite the $l=3$ inertial modes at lower frequencies, but in this paper we do not consider contributions to $\phi$ arising from $l>4$ spherical harmonics.

We also note that the set of frequencies emerging from the origin in figures \ref{fig:l3resonantOrbitalFreqs} and \ref{fig:l4resonantOrbitalFreqs}, representing the orbital frequencies at which a misaligned orbit may resonate with the $l=3$ and $l=4$ inertial modes, have previously been calculated in the slowly-rotating limit in \cite{Ogilvie2013}.
We agree with this calculation that, for a misaligned orbit in this limit, resonances with inertial waves occur for $l=3$ components of the potential at
\begin{align}
\frac{\omega_{\mathrm{orbit}}}{\Omega} = \pm \{ 0.1700, & 0.2981, 0.3922, 0.4444, 0.5099, \nonumber \\
& 0.8944, 1.1766, 1.3333 \} \, ,
\end{align}
and for $l=4$ components at
\begin{align}
\frac{\omega_{\mathrm{orbit}}}{\Omega} = \pm \{ 0.0970, & 0.1770, 0.1920, 0.1940, 0.3273, 0.3540,\nonumber \\
& 0.3840, 0.4550, 0.5580, 0.625, 0.6547, \nonumber \\
& 0.9100, 1.1160, 1.25 \} \, .
\end{align}
However, the claimed resonances of \cite{Ogilvie2013} for $\omega_{\mathrm{orbit}} = \pm 0.25 \Omega$ and $\pm 0.5 \Omega$ are both based on the existence of modes having $(l, m, \kappa) = (4, \pm 3, \mp 2)$.
On the basis of appendix D of paper I, we do not believe that such modes exist.

\section{The Power Dissipated due to a Misaligned Companion}
\label{sec:oblique_orbit_power}

In the previous section we described the gravitational potential produced by a point-mass on a circular orbit that is misaligned from the equatorial plane of the Maclaurin spheroid by an angle $\theta_i$.
In this section we shall consider the time-averaged rate of energy dissipation by the tide raised by this point-mass,
\begin{align}
\label{powerDef}
\left< P \right> & = \rho \int_V \left< \mathrm{Re}\left(\bm{u}\right) \cdot \mathrm{Re}\left(-\bm{\nabla}\phi\right) \right> \mathrm{d} V \,.
\end{align}
We chose to compute the power, rather than the torque, as a diagnostic for the influence of resonances on the orbital evolution owing to its simplicity.
The power is a positive definite scalar, whereas the torque is a vector.
Computing the $z$ component of the torque alone would still leave us with a signed quantity.

Were the time dependence of $\phi$ to consist of only a single Fourier mode $\propto e^{-i \kappa \Omega t}$, this time-averaged power would be
\begin{align}\label{PowerLambdaphiMainText}
& \left< P \right> =
\frac{c \kappa \Omega}{2 G}
\sum_{l=0}^{\infty} \sum_{m=-l}^l
\frac{(l+m)!}{(2l+1)(l-m)!} \nonumber \\
& \times \frac{B_l^m(\zeta_0)}{1 + \zeta_0 B_l^m(\zeta_0) \left(1 - \zeta_0 \cot^{-1} \zeta_0\right)}
\mathrm{Im} \left[\sum_{\kappa^{\prime}} D_{l,m,\kappa^{\prime}}^* \Lambda_{l,m,\kappa^{\prime}}^* \phi_{(o)}{}_l^m \right] \,.
\end{align}
We derive this (and explain the meaning of the notation) in appendix \ref{sec:power_torque_calculation}.
However, in our problem $\phi$ contains a myriad of frequencies, with $\phi_{(o)}{}_l^m \propto \sum_{m^{\prime}} \exp\left( -i \left( m^{\prime}\omega_{\mathrm{orbit}} - m \Omega \right) t \right)$.
In the interests of clarity, we introduce a new piece of notation, $\phi_{(o)}{}_l^m\left[ m^{\prime} \right]$, such that
\begin{align}
\label{phi[]def}
\phi = \sum_{l=0}^\infty \sum_{m=-l}^l \sum_{m^{\prime} = -\infty}^{+\infty} \Bigg( &
\phi_{(o)}{}_l^m \left[ m^{\prime} \right]
\frac{P_l^m\left(i \zeta\right) P_l^m\left(\mu\right)}{P_l^m\left(i\zeta_0\right)} \nonumber \\
\times e^{im\varphi} & \exp\left(-i \left(m^{\prime} \omega_{\mathrm{orbit}} - m \Omega \right)t \right) \Bigg) \,.
\end{align}
The reader might be concerned that, if $\omega_{\mathrm{orbit}} / \Omega$ is rational, the time dependence alone does not fix $m^{\prime}$ and $m$ separately;
however, the azimuthal dependence still fixes the value of $m$, so the symbol $\phi_{(o)}{}_l^m\left[ m^{\prime} \right]$ is still well-defined.
In appendix \ref{sec:power_torque_calculation} we give an expression for this symbol, and show that the average power may be expressed as
\begin{align}
\label{PExprMainText}
\left< P \right>
= \frac{c}{G} & \sum_{l=0}^{\infty} \sum_{m=-l}^l \sum_{m^{\prime} = -\infty}^{+\infty} \Bigg( \frac{\left(l+m\right)!}{\left(2l+1\right)\left(l-m\right)!} \nonumber \\
\times & \frac{B_l^m\left(\zeta_0\right) \left(m^{\prime}\omega_{\mathrm{orbit}} - m\Omega\right)}{1 + \zeta_0 B_l^m\left(\zeta_0\right)\left(1-\zeta_0\cot^{-1}\zeta_0\right)} \nonumber \\
\times & \mathrm{Im}\left[ \sum_{\kappa^{\prime}} \phi_{(o)}{}_l^m\left[m^{\prime}\right] D_{l,m,\kappa^{\prime}}^* \Lambda_{l,m,\kappa^{\prime}}^* \left[ m^{\prime} \right] \right]
\Bigg) \,.
\end{align}

Note that $\left< P \right>$ is \emph{not} the work done on the spheroid by the planet in the inertial frame, which could take either sign.
$\bm{u}$ is the velocity in the rotating frame (equivalently, the velocity perturbation relative to solid body rotation in the inertial frame), so \eqref{powerDef} is the work done on the fluid \emph{in the rotating frame}.
This quantity is always positive, and it may be shown from \eqref{LambdaForcingRelationMainText} that $\left< P \right>$ is the rate at which the fluid dissipates energy through viscosity,
\begin{align}
\label{PDisp}
\left< P \right> = 2 \rho \nu \sum_{\alpha, \beta, m^{\prime}} \int_V \bigg( & \left( \Lambda_{\beta} \left[ m^{\prime} \right] e_{\beta i j} \left[ m^{\prime} \right] \right)^* \nonumber \\
 & \times \left( \Lambda_{\alpha} \left[ m^{\prime} \right] e_{\alpha i j} \left[ m^{\prime} \right] \right) \mathrm{d} V \bigg) \,.
\end{align}
In the above sum, $\alpha$ and $\beta$ index modes of matching order, $m$, but possibly different degrees, $l$.
The square brackets notation was described earlier in this section;
$m^{\prime}$ in combination with $m$ fixes the forcing frequency.
We have checked that our numerical calculations of the power produced via \eqref{PExprMainText} agree with that computed via \eqref{PDisp}.

\begin{figure*}
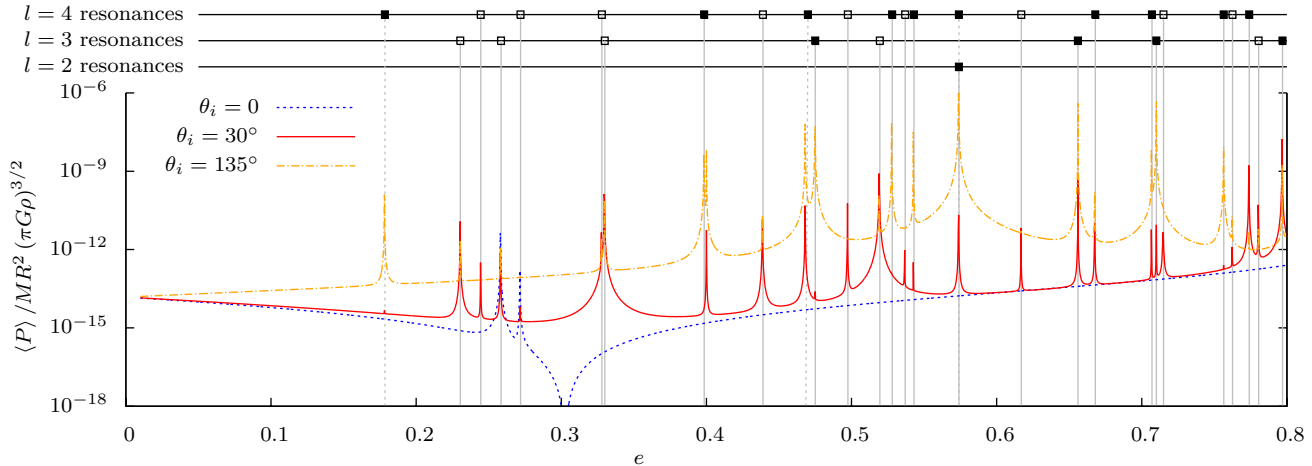

\centering
\include{figures/P_vs_e/R3/R3}
\caption{
Time averaged power dissipated in the tide raised by a point mass $10^{-3} M$ on a circular orbit of inclination $\theta_i$ relative to the equatorial plane of the spheroid.
The viscosity has been chosen such that the quality factor due to the $l=2$, $m=2$ tide is $Q = 10^7$ at $e=0$.
The orbital radius is $r_{\mathrm{orbit}} = 3 R$, where $R$ is the mean radius of the spheroid.
The horizontal axis shows the eccentricity of the Maclaurin spheroid, higher eccentricities corresponding to more rapid rotation rates.
For $e \approx 0.3025$ the rotation frequency of the spheroid matches the orbital frequency.
The vertical grey lines denote the locations at which resonances may be excited by an orbit with a generic misalignment.
An open box at the top of the line denotes that the resonance is with an inertial mode;
a solid box corresponds to a surface gravity mode.
A dashed vertical line indicates that this is a resonant excitation of an $l=2$ mode by the $l=4$ (spherical) component of the gravitational potential, via the transformation of equation \eqref{sphertooblatetrans}.
}
\label{fig:P_vs_e_R3}
\end{figure*}

\begin{figure*}
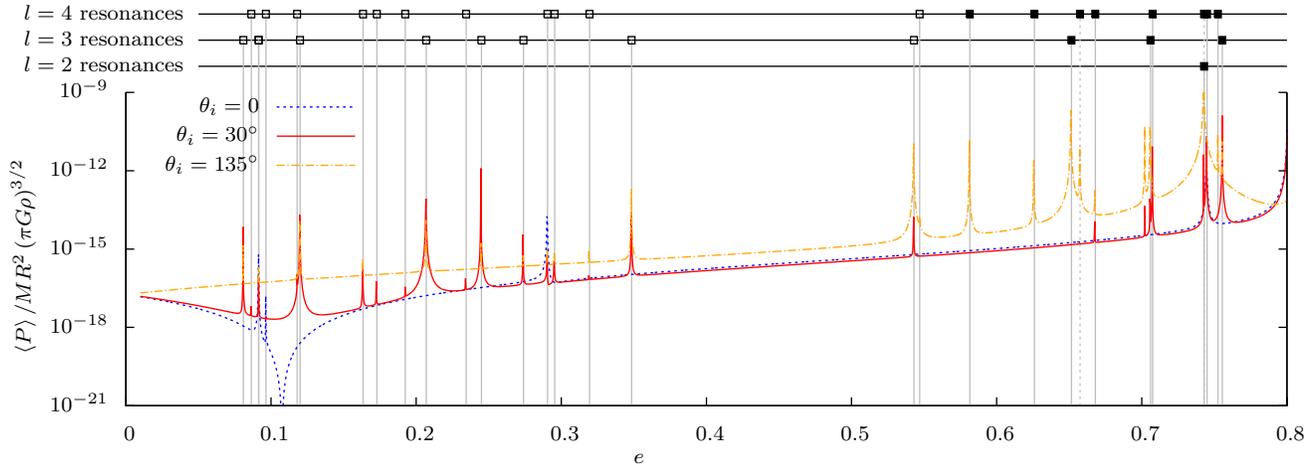

\centering
\include{figures/P_vs_e/R6/R6}
\caption{
As for figure \ref{fig:P_vs_e_R3}, but with the point-mass companion placed on an orbit of radius $r_{\mathrm{orbit}} = 6 R$.
The kinematic viscosity (as opposed to the quality factor at $e=0$) has been kept the same between these two figures.
For $e \approx 0.1075$ the rotation frequency of the spheroid matches the orbital frequency.
}
\label{fig:P_vs_e_R6}
\end{figure*}

In section \ref{sec:oblique_orbit_potential} we speculated that misaligned orbits would dissipate power at a greater rate, and we may now examine this claim by direct computation.
In figure \ref{fig:P_vs_e_R3} we plot the mean power dissipated for an orbital radius of three times the mean radius of the Maclaurin spheroid, versus the eccentricity of the spheroid.
Whilst extreme, such close orbits are not unheard of:
WASP-12b, WASP-103b, and Kepler-78b all have semi-major axes less than three times the radius of their host star, and there are eight more planets with $a / R_* < 4$.
The kinematic viscosity was chosen to be $\nu / R^2 \sqrt{\pi G \rho} = 2.25 \times 10^{-6}$.
This is the viscosity such that, in the slowly rotating limit, the $l=2$, $m=2$ tidal component has a quality factor of
\begin{align}
Q = 2 \pi \frac{E_0}{\Delta E}
= \omega_{\mathrm{orbit}} \frac{\epsilon_{\alpha}}{2 \nu \int_V \bm{e}_{\alpha} : \bm{e}_{\alpha} \mathrm{d} V}
= 10^7 \,.
\end{align}
In the above expression $E_0$ is the energy contained in the tide, $\Delta E$ is the energy dissipated in a single period, $\epsilon$ is defined in \eqref{epsilonDef} and is proportional to the energy of the mode, $\bm{e}$ is the rate-of-strain tensor defined just below equation \eqref{deltakappaDef}, and $\alpha$ is the index of the $l=2$, $m=2$ mode.

For much of figure \ref{fig:P_vs_e_R3} the response consists of a slowly varying `baseline' due to the $l=2$, $m=2$ surface gravity mode.
This corresponds to the equilibrium tide, and is forced by a large amplitude harmonic well below the natural frequency of the mode.
There is a very obvious feature around $e \approx 0.3$ where the tidal power generated by the aligned orbit drops sharply.
This is due to $e = 0.3025$ being the eccentricity at which the rotation period of the spheroid matches the orbital period of an orbit of $r_{\mathrm{orbit}} = 3 R$.
The potential due to the aligned orbit is then static in the frame rotating with the spheroid.
The orbit with a misalignment of $30^{\circ}$ shows noticeably greater tidal dissipation over the range $0.2 \lesssim e \lesssim 0.4$, since the potential due to such an orbit is not static in the rotating frame;
this orbit is continuing to excite $l=2$ surface gravity modes.
We also see that the misaligned orbit does experience a much richer set of resonances, as we predicted in section \ref{sec:oblique_orbit_potential}.
The baseline of the retrograde orbit ($\theta_i = 135^{\circ}$) shows markedly different behaviour than its prograde equivalents at low eccentricities.
This is due to the power associated with the $l=2$, $m=\pm2$ equilibrium tide scaling with the square of the forcing frequency. 
The frequency with which this tide is being forced increases with the rotation rate of the spheroid for a retrograde orbit, in contrast to the prograde case for $e \lesssim 0.3$.
The retrograde orbit also experiences enhanced dissipation due to several resonances, but many of the peaks seen in the $\theta_i = 30^{\circ}$ case are masked by the greater dissipation from the equilibrium tide.

We have also marked on figure \ref{fig:P_vs_e_R3} the predicted spheroid eccentricities at which gravitational spherical harmonics of degrees $l=2$, $3$ and $4$ could resonantly excite modes.
We have indicted which gravitational harmonic these originate from, and whether the mode excited is a surface gravity mode or an inertial mode.
For the aligned orbit, the only visible resonances are with an $l=3$ and an $l=4$ inertial mode.
For the orbit inclined by $\theta_i = 30^{\circ}$ many more resonances with inertial modes are visible.
In addition, at $e \approx 0.3985$, there is a clear resonance with an $l=4$ surface gravity mode.
This is a comparable eccentricity to Jupiter ($e=0.354$) and Saturn ($e=0.432$);
were these planets to have possessed satellites in such a configuration, this resonance would have played a role in their orbital evolution.
It is conceivable that this resonance may be important for the orbit of Mimas, which has an inclination of $1.6^{\circ}$ relative to the equator of Saturn, and for which $a_{\mathrm{Mimas}} / R_{\mathrm{Saturn}} \approx 3.2$.
This is, however, a much lower inclination than that which we have considered here, and without computing the tidal torque we cannot say what the effect on the orbit of Mimas would be.

In figure \ref{fig:P_vs_e_R6} we repeat these calculations for an orbital radius of $6 R$.
In addition to the 39 confirmed exoplanets with $5.5 < a / R_* < 6.5$, this is also the ratio of the semi-major axis of Io to the radius of Jupiter.
The effects of misalignment that we remarked upon in the previous paragraph are still present, but are much less dramatic than at $r_{\mathrm{orbit}} = 3R$.
The $\theta_i = 30^{\circ}$ orbit still shows enhanced dissipation around corotation, $0.075 \lesssim e \lesssim 0.13$.
Although this range is shorter than that experienced by the closer orbit, it is at lower eccentricities which may be more typical for stars.
For rapidly spinning spheroids (e.g. $e \approx 0.4$) the additional dissipation experienced by the retrograde ($\theta_i = 135^{\circ}$) orbit is much less for $r_{\mathrm{orbit}} = 6R$ than we saw in the $r_{\mathrm{orbit}} = 3R$ case.
This is due to the lower orbital frequency, which is now less significant than the angular velocity of the spheroid.

The excitation of resonances for this orbital radius is much more sparse.
The resonances of the $l=4$ potential with inertial modes, when they are visible at all, are very small.
In addition, the resonances with surface gravity modes do not occur until $e \approx 0.5815$, a rather extreme eccentricity.

\begin{figure*}
\centering
\input{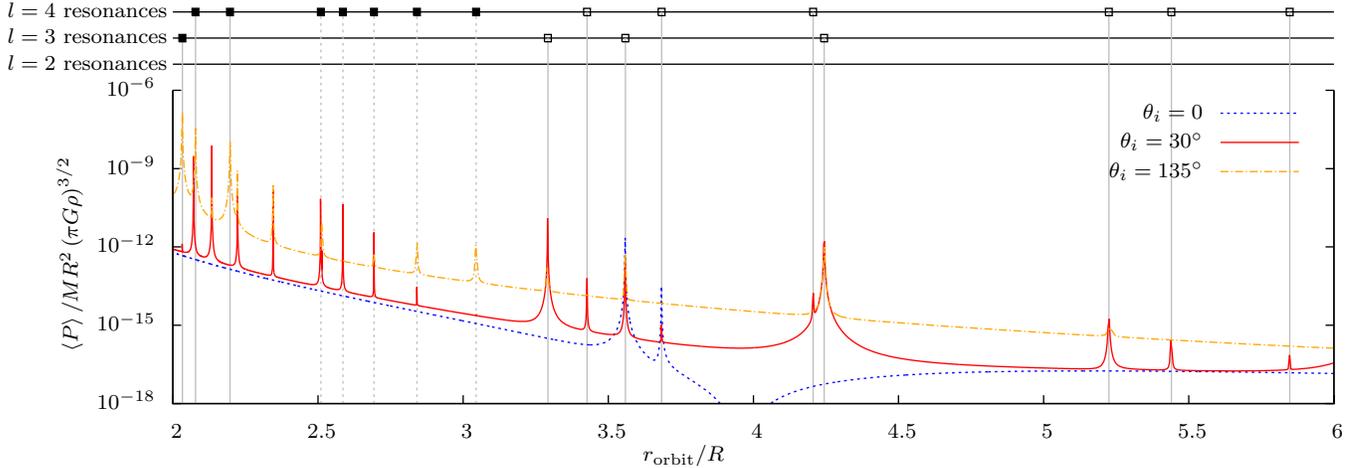}
\caption{
Time averaged power dissipated in the tide raised by a point mass $10^{-3} M$ on a circular orbit of inclination $\theta_i$ relative to the equatorial plane of the spheroid.
The spheroid is of eccentricity $e=0.2$, and its rotation period matches the orbital period at $r_{\mathrm{orbit}} = 3.96 R$.
The kinematic viscosity was taken to be the same as that of figures \ref{fig:P_vs_e_R3} and \ref{fig:P_vs_e_R6},  $\nu / R^2 \sqrt{\pi G \rho} = 2.25 \times 10^{-6}$.
}
\label{fig:P_vs_r_e02}
\end{figure*}

\begin{figure*}
\centering
\input{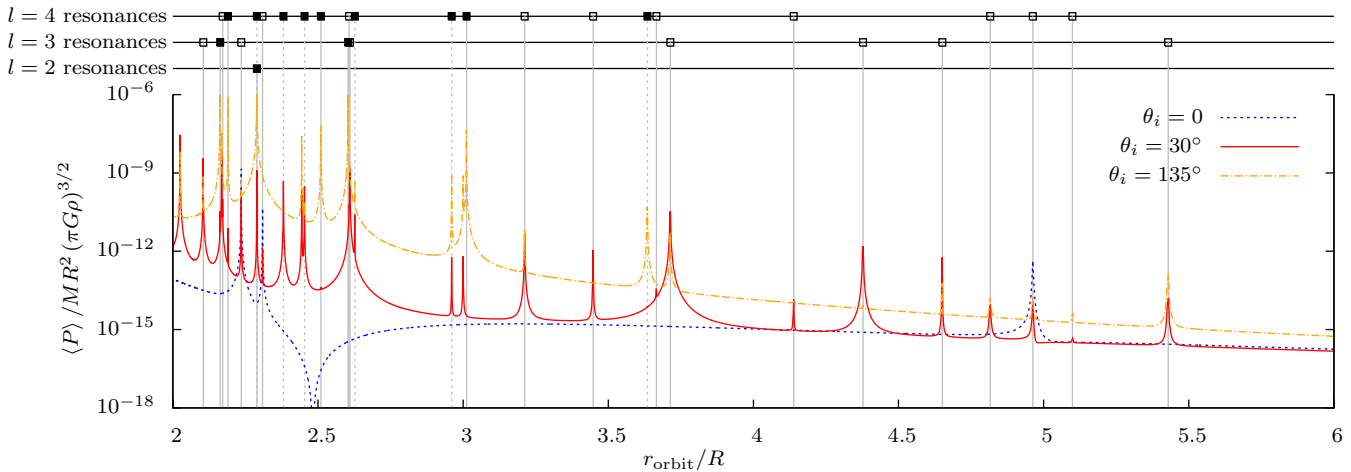}
\caption{
As for figure \ref{fig:P_vs_r_e02}, but for a spheroid of eccentricity $e=0.4$.
The rotation period of the spheroid matches the orbital period at $r_{\mathrm{orbit}} = 2.42 R$.
}
\label{fig:P_vs_r_e04}
\end{figure*}

Figures \ref{fig:P_vs_e_R3} and \ref{fig:P_vs_e_R6} may be thought as as showing the resonances that would be encountered by a planet on a fixed orbit as a young host-star spins down due to magnetic braking (moving from right to left along the $e$ axis).
A more typical scenario to consider would be a star rotating at a constant rate, with a planet moving inwards under the action of tidal dissipation.
In figure \ref{fig:P_vs_r_e02} we plot the power dissipated against orbital radius, for a star of eccentricity $e=0.2$, a planet of mass $M_2 = 10^{-3} M$, and a kinematic viscosity of $\nu / R^2 \sqrt{\pi G \rho} = 2.25 \times 10^{-6}$ (matching that used for the earlier plots).
We see that a planet in an equatorial orbit encounters very few resonances, whereas an inclination of $30^{\circ}$ results in very a large number of very sharp resonances (with peak dissipations enhanced by three orders of magnitude), but also a pair of broader resonances (the $l=3$ inertial modes excited at $r_{\mathrm{orbit}} = 3.29 R$ and $r_{\mathrm{orbit}} = 4.24 R$) that increase the dissipation by an order or magnitude over a significant range.
As seen earlier, the retrograde orbit has a consistently greater dissipation rate.

In figure \ref{fig:P_vs_r_e04} we perform the same experiment with a spheroid of eccentricity $e=0.4$.
This moves the inertial modes to higher frequencies, we no longer see the clean separation between resonances with inertial modes at large $r_{\mathrm{orbit}}$, and with surface gravity modes at small $r_{\mathrm{orbit}}$, that figure \ref{fig:P_vs_r_e02} exhibited. 
Inside of corotation, the $\theta_i = 30^{\circ}$ orbit causes tidal dissipation that is consistently an order of magnitude higher than the aligned case, a distinction that was not present for the $e=0.2$ spheroid.

\section{Approximating the response by potential Love numbers}
\label{sec:oblate_Love_numbers}

A common parameterisation of tidal response is the \emph{potential Love number}, which relates the potential due to the tidal deformation, $\Phi^{\prime}$, to the potential causing the deformation, $\phi$.
These are a function of the degree ($l$) and order ($m$) of the gravitational harmonic, and of the forcing frequency $\omega$, and are denoted by $k_l^m \left( \omega \right)$.
The relation between the gravitational forcing and the response is then
\begin{align}
\Phi^{\prime}
= \sum_{\omega} \sum_{l=2}^{\infty} \sum_{m=-l}^l k_l^m \left(\omega\right) \phi_{(s)}{}_l^m\left( \omega \right) Y_l^m \left( \theta, \varphi \right) e^{-i \omega t} \,.
\end{align}
The magnitude of $k_l^m$ encodes the amplitude of gravitational potential due to the tide raised, while its argument encodes the phase lag.
$\mathrm{Im} \left[ k_l^m \right]$ determines the dissipation.

Examination of equation \eqref{LambdaForcingRelationMainText} suggests that the Maclaurin spheroid model cannot strictly be described by such a parameterisation.
Firstly, we showed in appendix \ref{sec:power_torque_calculation} that the right hand side of \eqref{LambdaForcingRelationMainText} is proportional to a coefficient of an expansion of $\phi$ in oblate spheroidal harmonics, rather than spherical harmonics;
forcing by a spherical harmonic will not produce a single spherical harmonic as the gravitational response.
Secondly, the $ \sum_{\alpha} \int_V \bm{e}_{\beta}^* : \bm{e}_{\alpha} \mathrm{d} V$ term on the left hand side of \eqref{LambdaForcingRelationMainText} `mixes' terms of equal order, $m$, and whose degrees, $l$, have matching parity;
forcing by an oblate spheroidal harmonic will not produce a single oblate spheroidal harmonic as the gravitational response.

However, we conjectured in section \ref{sec:forced_problem} that this mixing integral between different modes would be small.
Inspection of the plots in paper I of the velocity fields shows very different structure between modes, with the exception of the prograde and retrograde components of the surface gravity modes.
Since we have already numerically evaluated solutions to \eqref{LambdaForcingRelationMainText} in section \ref{sec:oblique_orbit_power}, we can do the same for various levels of approximation to \eqref{LambdaForcingRelationMainText} and compare the results.

\begin{figure*}
\centering
\input{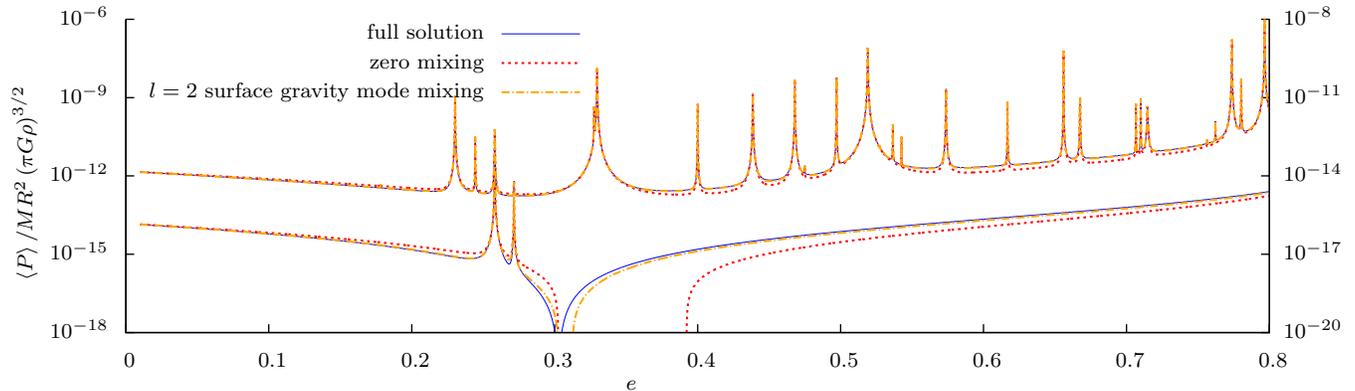}
\caption{
Repeat of the power calculations for the equatorial orbit, and for the orbit misaligned by $\theta_i = 30^{\circ}$ shown in figure \ref{fig:P_vs_e_R3}.
We show the full solution to equation \eqref{LambdaForcingRelationMainText} retaining modes with $l \le 4$ (solid blue curve, identical to curves shown in figure \ref{fig:P_vs_e_R3}), an approximation in which all viscous mixing between modes is neglected (dashed red curve), and an approximation in which only the mixing between $l=2$ surface gravity modes is retained (dot-dashed orange curve).
The power curves of the orbit with $\theta_i = 30^{\circ}$ have been displaced upwards for readability;
the scale on the left hand side refers to the $\theta_i = 0$ orbit, the scale on the right refers to the $\theta_i = 30^{\circ}$ orbit.
}
\label{fig:mixing_comp_R3}
\end{figure*}

Neglecting the mode mixing entirely gives
\begin{align}
\label{LambdaZeroMixing}
\Lambda_{l,m,\kappa^{\prime}} \left[ m^{\prime} \right]
= & \frac{c}{G \rho}
\frac{B_l^m\left( \zeta_0 \right)}{1 + \zeta_0 B_l^m\left( \zeta_0 \right) \left( 1- \zeta_0 \cot^{-1} \zeta_0\right)} \nonumber \\
& \times \frac{\left(l+m\right)!}{\left(2l+1\right)\left(l-m\right)!}
D_{l,m,\kappa^{\prime}}^* \phi_{(o)}{}_l^m \left[ m^{\prime} \right] \nonumber \\
& \times \frac{\kappa^{\prime} \Omega}{\epsilon_{l,m,\kappa^{\prime}} \Omega \left( \left(m^{\prime} \omega_{\mathrm{orbit}}/\Omega - m \right) - \kappa^{\prime} - \delta \kappa_{l,m,\kappa^{\prime}} \right)} \,.
\end{align}
The definitions of $B_l^m\left( \zeta_0\right)$, $\epsilon_{\alpha}$, and $\delta \lambda_{\alpha}$ are given in equations \eqref{BDef}, \eqref{epsilonDef}, and \eqref{viscousEvalPert} respectively.
We repeated our earlier calculation of the tidal power dissipated due to forcing from a $M_2 = 10^{-3} M$ mass in an $r_{\mathrm{orbit}} = 3R$ orbit around a Maclaurin spheroid with viscosity $\nu / R^2 \sqrt{\pi G \rho} = 2.25 \times 10^{-6}$ (shown in figure \ref{fig:P_vs_e_R3}) for the aligned orbit, and the misaligned prograde orbit.
We show the power curves produce by this zero-mixing approximation in figure \ref{fig:mixing_comp_R3} as the red dashed curve.
The agreement is excellent near resonances, since these are dominated by a single mode.
However, away from resonances a larger fractional error emerges, visible as an underestimation of the power for a spheroid rotating more rapidly than the orbital frequency, and an overestimation of the power otherwise.
This error becomes particularly extreme for the aligned orbit near corotation.

We conjecture that this error is largely due to the following facts:
away from resonances the dissipation is dominated by the off-resonant excitation of the $l=2$ surface gravity modes;
the prograde and retrograde surface gravity waves have very similar velocity and shear profiles, and hence the overlap integral between the shear tensors of these modes will be significant.
We therefore compute an alternative approximation in which we retain the mixing terms in \eqref{LambdaForcingRelationMainText} between the $l=2$ surface gravity modes.
The $\Lambda_{\beta}$ coefficients for all other modes are taken to be given by \eqref{LambdaZeroMixing}.
This approximation is plotted as the orange dot-dashed curve in figure \ref{fig:mixing_comp_R3}.
We see that, while there is still a substantial error for the equatorial orbit when the spheroid is rotating nearly synchronously with the orbit, the fractional error elsewhere has been reduced significantly.

This latter approximation does not involve viscous mixing between modes of different degrees.
Therefore, in this approximation, forcing the spheroid with a pure oblate spheroidal harmonic, $P_l^m\left( i \zeta \right) P_l^m \left( \mu \right) e^{i m \varphi}$, will produce an exterior oblate spheroidal harmonic of the degree and order, $Q_l^m\left( i \zeta \right) P_l^m \left( \mu \right) e^{i m \varphi}$, as the gravitational response.
Thus the response of the spheroid may indeed be well-approximated by a series of \emph{oblate potential Love numbers}, $k_{(o)}{}_l^m$, which will in general be frequency-dependent.

\section{Conclusion}
\label{sec:conclusion}

We have shown in section \ref{sec:forced_problem} how the problem of a tidally forced, uniformly rotating fluid body may be written as a self-adjoint problem, and how the solution of this problem may be sought as a decomposition in terms of the free modes described in paper I.
In section \ref{sec:oblique_orbit_potential} we described how to compute the potential due to a point-mass companion on an oblique, circular orbit.
The calculation of this potential in terms of spherical harmonics is not new;
however, our problem requires the potential as a sum of the oblate spheroidal harmonics used by \citet{Bryan1889}.
We have described the procedure to compute these components, and tabulated the required coefficients for degrees $l \le 4$ in appendix \ref{sec:oblate_spherical_harmonic_conversions}.
We noted that, in addition to the ability of an oblique orbit to excite a much greater set of modes, the conversion between the spherical and oblate spheroidal harmonics allows higher degree gravitational harmonics to excite lower degree modes, provided that the eccentricity of the Maclaurin spheroid is high.

In section \ref{sec:oblique_orbit_power} we computed the tidal power due to orbits of various obliquities at three and six times the mean radius of the spheroid.
These were plotted against the eccentricity.
The eccentricity of a young star due to rotational flattening will decrease as the star spins down due to magnetic braking, and the system may be considered to move from right to left on figures \ref{fig:P_vs_e_R3} and \ref{fig:P_vs_e_R6}.
We see that the period of low dissipation as the corotation radius passes the orbit of the planet will only be experienced for an equatorial orbit, and will be unnoticeable for a misalignment of $30^{\circ}$.
In addition, a great number of resonances with inertial modes will be encountered.
For $a / R_* = 6$, the only significant resonances are with $l=3$ inertial modes.
For an orbit as close as $a / R_* = 3$, resonances with $l=4$ inertial modes can also be significant.
In addition, the resonant excitation of either the $l=2$ or the $l=4$ surface gravity modes by the $l=4$ tidal potential can produce increases in dissipation of the same magnitude and width as those due to resonances with inertial modes.
All of these factors suggest that for short-period planets around rapidly rotating stars the $l=4$ tide can be significant.
The same may be true for short-period satellites around giant planets.

In section \ref{sec:oblate_Love_numbers} we showed that the apparently complex response of the spheroid to forcing, which involves viscous mixing between modes due to the summation term on the left hand side of \eqref{LambdaForcingRelationMainText}, may be considerably simplified.
Neglecting this mixing entirely produces a moderate error far from resonances, and a very significant error for an orbit that is nearly synchronous with the rotation of the spheroid.
However, by including the mixing between only the $l=2$ surface gravity modes this error is largely removed.
Since there is no mixing between modes of different orders, $m$, and this approximation removes all the mixing between modes of different degrees, $l$, then the response of the spheroid may be well approximated by a series of \emph{oblate potential Love numbers}, $k_{(o)}{}_l^m \left( \lambda \right)$.
We do not compute such Love numbers here, confining our calculations to the full power curves for a small selection of orbits, since we wish to emphasise both the role of the higher degree modes and that of that higher degree gravitational harmonics.
However, we recognise that the potential Love numbers are a widely used prescription for the tidal response, and the purpose of section \ref{sec:oblate_Love_numbers} is to show that this prescription may still be used, albeit for a different set of harmonics from those usually considered.

While the work of this paper has shown that resonance with both the inertial and surface gravity modes may significantly contribute to the tidal dissipation in the host star of short period planets, and that this dissipation could be significantly higher for a misaligned orbit, we have not truly examined the effect on the orbital evolution.
To do so we must also compute the torque.
These results would then form the input for a calculation of the orbital evolution.
We might find more rapid inward migration for misaligned planets;
alternatively, the tide might act to rapidly damp the inclination of the planet relative to the equatorial plane of their host star, with little additional migration.
We intend to perform such computations in a subsequent paper.

An additional limitation of this work is that all of our calculations have been linear.
Figures \ref{fig:P_vs_e_R3} to \ref{fig:P_vs_r_e04} show resonances in which the tidal power is enhanced by up to three orders of magnitude.
These might be poorly described by linear theory.
The modes may break in such a manner as to set up differential rotation within the star or planet, an effect that was observed by \citet{Favier2014} in numerical simulations of a rotating fluid body with a solid.
The picture here of sharp resonances that a migrating planet or satellite passes through quickly may thus be broken by non-linear effects.
Investigating these effects would require numerical simulations of the full Navier-Stokes equations.
If convection were included, the small size of a convection cell relative to the star or planet would limit the simulations to a local patch of the body.

\section*{Acknowledgements}

This research was supported by the STFC. Harry Braviner is partially supported by Trinity College.

\appendix

\section{Decomposition of the forced solution}
\label{sec:decomposition}

In paper I we described the free modes of an inviscid Maclaurin spheroid, originally found by \citet{Bryan1889}, following the more modern treatment of \citet{LI1999}.
We also found the decay rates, to first order in viscosity, for the modes of a Maclaurin spheroid composed of a fluid of low viscosity.
Now we wish to derive the flow in a Maclaurin spheroid subject to tidal forcing, and it is natural to seek to write this solution as a sum over the inviscid modes whose properties we are already familiar with.
In paper 1 we introduced a compact notion 
for the inviscid modes, writing
\begin{align}
\lambda_{\alpha} = \kappa_{\alpha} \Omega, \;\;\;\;
& \bm{x}_{\alpha} = \left( \begin{array}{c} \bm{\xi}_{\alpha} \\ \bm{u}_{\alpha} \end{array} \right), \\
\bm{\mathrm{M}} = \left( \begin{array}{cc} \mathrm{C} & 0 \\ 0 & 1 \end{array} \right), \;\;\;\;
& \bm{\mathrm{K}}_{0} = \left( \begin{array}{cc} 0 & i\mathrm{C} \\ -i\mathrm{C} & -2 i \Omega \times \end{array} \right) \,.
\end{align}
The operator $\mathrm{C}$ is defined through its action on the displacement vector, $\mathrm{C}\bm{\xi} = \bm{\nabla} W$.
We have dropped the subscript `0' from paper I:
a quantity with a Greek index may be assumed to be that of a free, inviscid mode, and $i \delta \lambda_{\alpha}$ its decay rate.
In this notation the free, inviscid problem \eqref{inviscidNS} can be written as the eigenvalue problem
\begin{align} \label{inviscidNSmatrix}
\lambda_{\alpha} \bm{\mathrm{M}} \bm{x}_{\alpha} = \bm{\mathrm{K}}_0 \bm{x}_{\alpha} \,.
\end{align}
Using the fact that $\mathrm{K}_0$ is Hermitian we may find an orthogonality relation between these modes,
\begin{align} \label{matrixOrthogonality}
\left(\lambda_{\alpha} - \lambda_{\beta}^*\right) \int_V \bm{x}_{\beta}^{\dagger} \bm{\mathrm{M}} \bm{x}_{\alpha} \mathrm{d}V = 0 \,.
\end{align}
We made use of this in paper 1 to derive the decay rates of these modes when a small viscosity is introduced, and we shall make use of it again below.

Using $\delta$ to denote perturbations due to a small viscosity, we introduce the operator 
\begin{align}
\delta \bm{\mathrm{K}} = \left( \begin{array}{cc} 0 & 0 \\ 0 & i \nu \nabla^2 \end{array} \right)
\end{align}
and write the problem \eqref{viscousNS} for the free viscous modes as
\begin{align}
\left(\lambda_{\alpha} + \delta \lambda_{\alpha}\right) \bm{\mathrm{M}} \left(\bm{x}_{\alpha} + \bm{\delta x}_{\alpha}\right) 
= \left(\bm{\mathrm{K}}_0 + \delta\bm{\mathrm{K}}\right) \left(\bm{x}_{\alpha} + \bm{\delta x}_{\alpha}\right) \,.
\end{align}
In paper 1 we showed that, on the assumption that $\delta \bm{x}_{\alpha}$ is small everywhere except a thin free-slip boundary layer, the perturbation to the eigenvalue is given by
\begin{align} \label{viscousEvalPert}
\delta \lambda_{\alpha} = - 2 i \nu \frac{ \int_V e_{\alpha i j}^* e_{\alpha i j} \mathrm{d}V }
{ \int_V \bm{x}_{\alpha}^{\dagger} \mathrm{M} \bm{x}_{\alpha} \mathrm{d}V } \,.
\end{align}
We shall make use of this quantity shortly.

We now wish to solve \eqref{viscousforcedNS}, the forced, viscous problem.
After a sufficiently long time, we expect any transient motion to have decayed leaving a response which varies harmonically with the same frequency as the forcing, and we seek a decomposition of the spatial form of the response in terms of the free modes, as in \eqref{LambdaExpDef}.
Writing this solution as a sum over the inviscid, rather than viscous, modes will allow us to take advantage of the orthogonality relation \eqref{matrixOrthogonality}.
Defining $\bm{F} = \left( 0,\;\bm{f} \right)^{\mathrm{T}}$ we may the write the equation of motion as
\begin{align}
\lambda \sum_{\alpha} \Lambda_{\alpha} \bm{\mathrm{M}} \bm{x}_{\alpha}
= \sum_{\alpha} \Lambda_{\alpha} \bm{\mathrm{K}}_0 \bm{x}_{\alpha} + \sum_{\alpha} \Lambda_{\alpha} \delta \bm{\mathrm{K}} \bm{x}_{\alpha} + i \bm{F} \,.
\end{align}
Subtracting $\sum_{\alpha} \left( \Lambda_{\alpha} \eqref{inviscidNSmatrix} \right)$, left-multiplying by $\bm{x}_{\beta}^{\dagger}$ and integrating over the volume of the spheroid gives
\begin{align}
& \sum_{\alpha} \Lambda_{\alpha} \left(\lambda - \lambda_{\alpha}\right) \int_V \bm{x}_{\beta}^{\dagger} \bm{\mathrm{M}} \bm{x}_{\alpha} \mathrm{d}V \nonumber \\
= & \sum_{\alpha} \Lambda_{\alpha} \int_V \bm{x}_{\beta}^{\dagger} \delta \bm{\mathrm{K}} \bm{x}_{\alpha} \mathrm{d}V
+ i \int_V \bm{x}_{\beta}^{\dagger} \bm{F} \mathrm{d}V \,.
\end{align}
Some manipulation of $\delta \bm{\mathrm{K}}$ puts this into a more useful form.
\begin{align} \label{deltaKrewrite}
&\sum_{\alpha} \Lambda_{\alpha} \int_V \bm{x}_{\beta}^{\dagger} \delta \bm{\mathrm{K}} \bm{x}_{\alpha} \mathrm{d}V \nonumber \\
= & \sum_{\alpha} \Lambda_{\alpha} i \nu \int_V u_{\beta i}^* \partial_j \partial_j u_{\alpha i} \mathrm{d}V \nonumber \\
= & \sum_{\alpha} i \nu \Lambda_{\alpha} \int_{\partial V} n_j u_{\beta i}^* \partial_j u_{\alpha i} \mathrm{d}S
- \sum_{\alpha} i \nu \Lambda_{\alpha} \int_V \partial_j u_{\beta i}^* \partial_j u_{\alpha i} \mathrm{d}V \nonumber \\
= & 2 i \nu \int_{\partial V} n_j u_{\beta i}^* \left[ \sum_{\alpha} \Lambda_{\alpha} e_{\alpha i j} \right] \mathrm{d}S
- 2 i \nu \sum_{\alpha} \Lambda_{\alpha} \int_V e_{\beta i j}^* e_{\alpha i j} \mathrm{d}V \,.
\end{align}
In going from the third to the fourth line incompressibility has been used to recast derivatives of velocity into rate-of-strain tensors, $e_{\alpha i j} = \frac{1}{2} \left( \partial_i u_{\alpha j} + \partial_j u_{\alpha i} \right)$
By linearity, the quantity in square brackets in the final line is the rate-of-strain tensor of the viscous, forced solution.
We take this surface integral to vanish as our stress-free boundary condition.
Our basis of modes truncated at $l=4$ will not allow us to resolve the boundary layer itself, and we commented in the main text that this will not significantly affect the rate-of-strain tensor.
Therefore our dissipation rates will be accurate.
This should be contrasted to what we would see if we did not perform the above integration by parts and left the dissipation term as $\nu \bm{u} \cdot \nabla^2 \bm{u}$.
\cite{Zhang2004} found that $\int_V \bm{u} \cdot \nabla^2 \bm{u} \, \mathrm{d} V$ vanishes for each of the inviscid modes.
We note that this is consistent with our argument:
$\bm{\nabla} \bm{u}$ is $\mathcal{O} \left( u / R \right)$ in both the boundary and the interior, so the integral of this quantity in the free-slip boundary layer of thickness $\delta$ is $\mathcal{O}\left( \delta / R\right)$ relative to the contribution from the interior;
the requirement that some components of $\bm{\nabla} \bm{u}$ adjust to zero at the surface implies that $\nabla^2 \bm{u}$ is $\mathcal{O} \left( u / R \delta\right)$ in the boundary layer, whereas this quantity is $\mathcal{O}\left( u / R^2 \right)$ in the interior, allowing the volume integral $\int_V \bm{u} \cdot \nabla^2 \bm{u} \, \mathrm{d} V$ to vanish.

Using \eqref{matrixOrthogonality}, \eqref{viscousEvalPert} and \eqref{deltaKrewrite} we may write
\begin{align} \label{LambdaForcingRelation}
& \left(\lambda - \lambda_{\beta} - \delta \lambda_{\beta} \right) \epsilon_{\beta} \Lambda_{\beta} 
+ 2 i \nu \sum_{\alpha \ne \beta} \left( \int_V e_{\beta i j}^* e_{\alpha i j} \mathrm{d}V \right) \Lambda_{\alpha} \nonumber \\
= & i \int_V \bm{x}_{\beta}^{\dagger} \bm{F} \mathrm{d}V
= i \int_V \bm{u}_{\beta}^* \cdot \bm{f} \mathrm{d}V \,.
\end{align}
We have used the short-hand
\begin{align}
\epsilon_{\beta}
& = \int_V \bm{x}_{\beta}^{\dagger} \bm{\mathrm{M}} \bm{x}_{\beta} \mathrm{d}V \nonumber \\
& = \int_V \bm{u}_{\beta}^* \cdot \bm{u}_{\beta} \mathrm{d}V + \int_V \bm{\xi}_{\beta}^* \cdot \bm{\nabla} W_{\beta} \mathrm{d}V \,,
\end{align}
which is proportional to the energy, for the normalisation of the modes.


\section{Power and Torque due to an Orbiting Companion}
\label{sec:power_torque_calculation}

In the previous appendix we found that the expansion coefficients, $\Lambda_{\alpha}$, satisfied the relation \eqref{LambdaForcingRelation}.
We assumed that the forcing consisted of only a single harmonic, which is not the case for potentials we derived in section \ref{sec:oblique_orbit_potential}.
However, the fact that this is a linear theory allows us to compute the $\Lambda_{l,m,\kappa^{\prime}}$ coefficients separately for each forcing and sum them to get the total velocity field.

First let us consider a single Fourier mode of the forcing, $ \bm{f}e^{-i\kappa\Omega t} = - \bm{\nabla} \phi $, and write
\begin{equation} \label{phiOblateExpansion}
\phi = \sum_{l=0}^{\infty} \sum_{m=-l}^l \sum_{m^{\prime} = -\infty}^{+\infty} \widehat{\phi}_{(o)}{}_l^m \frac{P_l^m(\mu) P_l^m(i\zeta)}{P_l^m(i\zeta_0)} e^{i m \varphi - i \kappa \Omega t} \,.
\end{equation}
$\widehat{\phi}_{(o)}{}_l^m$ is a complex number;
it contains no time or space dependence.
Using incompressibility and the pressure boundary condition \eqref{pbc} we may write
\begin{align}
\label{pbc_force_trans}
i & \int_V \bm{u}_{\beta}^* \cdot \bm{f} \mathrm{d} V
= \kappa \Omega \int_{\partial V} \left(\bm{\xi}_{\beta} \cdot \bm{n}\right)^* \phi \mathrm{d} S \nonumber \\
& = \kappa \Omega \int_{\partial V} \left( \frac{-\rho \left(W - \Phi^{\prime}\right)_{\beta}^*}{\bm{n}\cdot\bm{\nabla}p} \right) \phi \mathrm{d} S \,.
\end{align}
\cite{LI1999} give the surface normal pressure gradient as
\begin{align}
\bm{n} \cdot \bm{\nabla} p = - 4 \pi c G \rho^2 \zeta_0 \sqrt{1 + \zeta_0^2} \left(1 - \zeta_0 \cot^{-1} \zeta_0\right) \sqrt{\zeta_0^2 + \mu^2}
\end{align}
and the area element on the surface is $\mathrm{d}S = c^2 \sqrt{1 + \zeta_0^2} \sqrt{\zeta_0^2 + \mu^2} \mathrm{d} \mu \mathrm{d} \varphi$.
Using the results of paper I and the expansion \eqref{LambdaExpDef} we may write
\begin{align}\label{WPhiFree}
\left(W - \Phi^{\prime}\right)_{\beta} = & \frac{\zeta_0 B_l^m(\zeta_0)\left(1 - \zeta_0 \cot^{-1}\zeta_0\right) D_{l, m, \kappa^{\prime}} }{1 + \zeta_0 B_l^m(\zeta_0)\left(1 - \zeta_0 \cot^{-1}\zeta_0\right)} \nonumber \\
& \times P_l^m(\mu) e^{i m \varphi - i \kappa \Omega t}
\end{align}
on the surface of the spheroid, where we have now chosen to expand the abstract mode index $\beta$ into the degree, $l$, order, $m$, and frequency $\kappa^{\prime} \Omega$, of the mode.
For each $l$ and $m$ there is a set of allowed values of $\kappa^{\prime}$ which satisfy the boundary conditions, as described and computed in paper I.
$D_{l, m, \kappa^{\prime}}$ is an arbitrary normalisation constant; it defines the magnitude of the velocity, displacement, and potential of a free mode, and will always appear in combination with the $\Lambda_{l,m,\kappa^{\prime}}$ coefficients.
$B_l^m(\zeta_0)$ is defined to be
\begin{align}\label{BDef}
B_l^m(\zeta_0) = & (1+\zeta_0^2) \nonumber \\
& \times \left.\left( \frac{1}{Q_l^m(i\zeta_0)} \frac{\mathrm{d}Q_l^m(i\zeta)}{\mathrm{d}\zeta} - \frac{1}{P_l^m(i\zeta_0)} \frac{\mathrm{d}P_l^m(i\zeta)}{\mathrm{d}\zeta}\right) \right|_{\zeta=\zeta_0} \,.
\end{align}
It originates from the surface boundary condition.

Combining these we may write
\begin{align} \label{freeModeGravInt}
& i \int_V \bm{u}_{\beta}^* \cdot \bm{f} \mathrm{d}V \nonumber \\
& = \frac{c}{G \rho} \frac{\kappa_{\alpha} \Omega B_l^m(\zeta_0) D_{l,m,\kappa^{\prime}}^* \widehat{\phi}_{(o)}{}_l^m}{1 + \zeta_0 B_l^m(\zeta_0)\left(1 - \zeta_0 \cot^{-1} \zeta_0\right)} \frac{(l+m)!}{(2l+1)(l-m)!} \,,
\end{align}
where it should be emphasised that $l$, $m$ and $\kappa^{\prime} \Omega$ are the degree, order and frequency of the free mode indexed by $\beta$, whereas $\kappa \Omega$ is the frequency of the forcing.

Due to the way we have computed $\phi$ in section \ref{sec:oblique_orbit_potential}, we have a slight complication when calculating the average power,
\begin{align}
\label{<P>def}
\left< P \right>
& = \int_V \rho \mathrm{Re}\left(\bm{u}\right) \cdot \mathrm{Re}\left(-\bm{\nabla} \phi \right) \mathrm{d} V \,.
\end{align}
We chose coefficients $\phi_{(o)}{}_l^m$ such that $\phi$ is real, rather than including only positive-frequency Fourier modes.
The price we pay is that we must consider cross-terms between modes of positive and negative frequency.
To make this discussion more precise, we defined $\phi_{(o)}{}_l^m \left[ m^{\prime} \right]$ in equation \eqref{phi[]def}.
The calculations of section \ref{sec:oblique_orbit_potential} may be used to show that
\begin{align}
\phi_{(o)}{}_l^m\left[m^{\prime}\right] &
= (-1)^{m+m^{\prime}+1} \frac{G M_2}{r_{\mathrm{orbit}}} P_l^m\left(i\zeta_0\right)
P_l^{m^{\prime}}\left( 0 \right)
\nonumber \\
& \times \hspace{-0.5cm} \sum_{\substack{l^{\prime} = \\ \max (l, |m^{\prime}| )}}^{\infty} \Bigg[ \sqrt{\frac{(l^{\prime}-m^{\prime})! (l^{\prime}-m)!}{(l^{\prime}+m^{\prime})! (l^{\prime}+m)!}}
\left(\frac{c}{r_{\mathrm{orbit}}}\right)^{l^{\prime}} \nonumber \\
& \hspace{1.5cm} \times \mathrm{H}^{(m)-1}_{l^{\prime} l}
d^{(l^{\prime})}_{m m^{\prime}} \left(\theta_i\right)
\Bigg] \,.
\end{align}
This should be interpreted as zero when $\left| m \right| > l$.
Note that $\mathrm{H}^{(m)-1}_{l^{\prime} l}$ is the $l^{\prime}$, $l$ component of the matrix inverse of $\bm{\mathrm{H}}^{(m)}$.

Write $\left\{ \Lambda_{l,m,\kappa^{\prime}} \left[ m^{\prime} \right] \;\; \left| \; l=0,1,2,\dots \right. \right\}$ to denote the set of coefficients found by solving \eqref{LambdaForcingRelationMainText} with the set $\left\{ \widehat{\phi}_{(o)}{}_l^m = \phi_{(o)}{}_l^m \left[ m^{\prime} \right] \;\; \left| \; l=0,1,2,\dots \right. \right\}$ of gravitational coefficients and with forcing frequency $\kappa \Omega = \left( m^{\prime} \omega_{\mathrm{orbit}} - m \Omega \right)$.
Note that we may consider the different orders ($m$) separately when solving the linear equation for the coefficients, but we may not do the same with the different degrees ($l$);
$\widehat{\phi}_{(o)}{}_l^m$ will produce contributions to $\Lambda_{l-2,m,\kappa^{\prime}}$, $\Lambda_{l+2,m,\kappa^{\prime}}$, etc.

The azimuthal integral in \eqref{<P>def} gives non-zero contributions when the orders, $m$, of the gravitational and response coefficients are either identical, or differ by a sign.
In the former case the values of $m^{\prime}$ must match, and in the latter they must also differ by a sign (else the time averaged contribution will vanish).
The calculation is essentially the same as that taking us from \eqref{pbc_force_trans} to \eqref{freeModeGravInt}, and we find that
\begin{align}
\label{<P>withcrossterms}
\left< P \right>
= \frac{c}{2G} \sum_{l=0}^{\infty} & \sum_{m=-l}^l \sum_{m^{\prime} = -\infty}^{+\infty} \Bigg(
\frac{1}{\left(2l+1\right)} \nonumber \\
\times & \frac{B_l^m\left(\zeta_0\right) \left(m^{\prime}\omega_{\mathrm{orbit}} - m\Omega\right)}{1 + \zeta_0 B_l^m\left(\zeta_0\right)\left(1-\zeta_0\cot^{-1}\zeta_0\right)} \nonumber \\
\times \sum_{\kappa^{\prime}} &
\mathrm{Im}\bigg[
(-1)^m \phi_{(o)}{}_l^m\left[m^{\prime}\right] D_{l,-m,\kappa^{\prime}} \Lambda_{l,-m,\kappa^{\prime}} \left[-m^{\prime}\right] \nonumber \\
& \;\; + \frac{(l+m)!}{(l-m)!} \phi_{(o)}{}_l^m\left[m^{\prime}\right] D_{l,m,\kappa^{\prime}}^* \Lambda_{l,m,\kappa^{\prime}}^* \left[m^{\prime}\right]
\bigg]
\Bigg) \,.
\end{align}
We can simplify this further.
Our $\phi$ is real by construction, which manifests itself as the relation
\begin{align}
\phi_{(o)}{}_l^m\left[ m^{\prime} \right]
= (-1)^m \frac{(l-m)!}{(l+m)!} \left( \phi_{(o)}{}_l^{-m}\left[-m^{\prime}\right]\right)^* \,.
\end{align}
Combining this with \eqref{<P>withcrossterms} and using the fact that $B_l^{-m}\left(\zeta_0\right) = B_l^m\left(\zeta_0\right)$ we may write
\begin{align}
\label{<P>realphi}
\left< P \right>
= \frac{c}{G} \sum_{l=0}^{\infty} & \sum_{m=-l}^l \sum_{m^{\prime} = -\infty}^{+\infty} \Bigg(
\frac{(l+m)!}{\left(2l+1\right)(l-m)!} \nonumber \\
\times & \frac{B_l^m\left(\zeta_0\right) \left(m^{\prime}\omega_{\mathrm{orbit}} - m\Omega\right)}{1 + \zeta_0 B_l^m\left(\zeta_0\right)\left(1-\zeta_0\cot^{-1}\zeta_0\right)} \nonumber \\
\times & \sum_{\kappa^{\prime}}
\mathrm{Im}\bigg[
\phi_{(o)}{}_l^m\left[m^{\prime}\right] D_{l,m,\kappa^{\prime}}^* \Lambda_{l,m,\kappa^{\prime}}^* \left[m^{\prime}\right]
\bigg]
\Bigg) \,.
\end{align}

For completeness we note that if $\phi$ is considered to be the real part of a single Fourier mode of frequency $\kappa \Omega$ and complex amplitudes $\widehat{\phi}_{(o)}{}_l^m$, the equivalent expression is
\begin{align}\label{PowerLambdaphi}
\left< P \right> =
\frac{c}{2 G}
\sum_{l=0}^{\infty} & \sum_{m=-l}^l
\frac{(l+m)!}{(2l+1)(l-m)!} \nonumber \\
& \times \frac{B_l^m(\zeta_0) \kappa \Omega }{1 + \zeta_0 B_l^m(\zeta_0) \left(1 - \zeta_0 \cot^{-1} \zeta_0\right)} \nonumber \\
& \times \mathrm{Im} \left[\sum_{\kappa^{\prime}} D_{l,m,\kappa^{\prime}}^* \Lambda_{l,m,\kappa^{\prime}}^* \widehat{\phi}_{(o)}{}_l^m \right]
\,.
\end{align}



\section{The Relation Between Oblate and Spherical Harmonics}
\label{sec:oblate_spherical_harmonic_conversions}

In \citet{Bryan1889} it was shown that the functions $P_l^m(i \zeta) P_l^m(\mu) e^{i m \varphi}$, which we refer to as \emph{interior oblate spheroidal harmonics}, solve Laplace's equation.
From this, their $\varphi$ dependence, and their regularity at the origin, we can deduce that they must be expressible as a sum over interior spherical harmonics of the same order, but possibly differing degree.
The most direct way to obtain such expansion is to convert $P_l^m(i \zeta) P_l^m( \mu)$ into cylindrical, and then spherical, coordinates using the definition \eqref{oblateSpherCoordDef}.
Denote the expansion coefficients by $\mathrm{H}^{(m)}_{l l^{\prime}}$, defined by
\begin{align}
P_l^m(i \zeta) P_l^m(\mu) = \sum_{l^{\prime} = 0}^{\infty} \mathrm{H}^{(m)}_{l l^{\prime}} \left(r / c\right)^{l^{\prime}} P_{l^{\prime}}^m \left(\cos \theta\right) \,.
\end{align}
These expansions turn out to be finite, and $\mathrm{H}^{(m)}_{l l^{\prime}} = 0$ for $l^{\prime} > l$.
We now list all non-zero $\mathrm{H}^{(m)}_{l l^{\prime}}$ for $l \le 4$:
\begin{align}
\mathrm{H}^{(0)}_{0 0} &= 1,& \;
\mathrm{H}^{(0)}_{1 1} &= i,& \;
\mathrm{H}^{(0)}_{2 0} &= -\frac{1}{2},& \; \mathrm{H}^{(0)}_{2 2} &= - \frac{3}{2}, \nonumber \\
\mathrm{H}^{(0)}_{3 1} &= -\frac{3 i}{2},& \; \mathrm{H}^{(0)}_{3 3} &= -\frac{5 i}{2}, && && \nonumber \\
\mathrm{H}^{(0)}_{4 0} &= \frac{3}{8},& \; \mathrm{H}^{(0)}_{4 2} &= \frac{15}{4},& \; \mathrm{H}^{(0)}_{4 4} &= \frac{35}{8}, && \nonumber \\
\mathrm{H}^{(1)}_{1 1} &= -1,& \;
\mathrm{H}^{(1)}_{2 2} &= -3i,& \;
\mathrm{H}^{(1)}_{3 1} &= 9,& \; \mathrm{H}^{(1)}_{3 3} &= \frac{15}{2}, \nonumber \\
\mathrm{H}^{(1)}_{4 2} &= 25i,& \; \mathrm{H}^{(1)}_{4 4} &= \frac{35 i}{2}, && && \nonumber \\
\mathrm{H}^{(2)}_{2 2} &= 3,& \;
\mathrm{H}^{(2)}_{3 3} &= 15i,& \;
\mathrm{H}^{(2)}_{4 2} &= -\frac{225}{2},& \; \mathrm{H}^{(2)}_{4 4} &= -\frac{105}{2}, \nonumber \\
\mathrm{H}^{(3)}_{3 3} &= -15,& \;
\mathrm{H}^{(3)}_{4 4} &= - 105i, && && \nonumber \\
\mathrm{H}^{(4)}_{4 4} &= 105 \,. && && &&
\end{align}
Note that expansions for negative $m$ may be obtained from the formula
\begin{align}
\mathrm{H}^{(-m)}_{l l^{\prime}} = (-1)^m \frac{(l^{\prime} + m)!}{(l^{\prime} - m)!} \left(\frac{(l - m)!}{(l + m)!}\right)^2 \mathrm{H}^{(m)}_{l l^{\prime}} \,,
\end{align}
and that in this paper we use the Condon-Shortley phase convention.

In our calculations, the quantities we actually make use of are the matrix inverses, $\bm{\mathrm{H}}^{(m)-1}$.
With the exception of $m=0$, the $\bm{\mathrm{H}}^{(m)}$ matrices are in fact singular and do no possess inverses.
What we mean by this slight abuse of notation is that $\sum_k \mathrm{H}^{(m)-1}_{lk} \mathrm{H}^{(m)}_{kl^{\prime}} = \delta_{l l^{\prime}}$ for $\left|m\right| \le l, l^{\prime}$, and hence that
\begin{align}
\left(\frac{r}{c}\right)^l P_l^m\left(\cos \theta \right) = \sum_{l^{\prime} = 0}^{\infty} \mathrm{H}^{(m)-1}_{l l^{\prime}} P_{l^{\prime}}^m\left(i\zeta\right) P_{l^{\prime}}^m\left(\mu\right) \,.
\end{align}
For $l \le 4$, the non-zero $\mathrm{H}^{(m)-1}_{l l^{\prime}}$ are
\begin{align}
\mathrm{H}^{(0)-1}_{0 0} &= 1,& 
\mathrm{H}^{(0)-1}_{1 1} &= -i,& 
\mathrm{H}^{(0)-1}_{2 0} &= -\frac{1}{3}, \nonumber \\ 
\mathrm{H}^{(0)-1}_{2 2} &= -\frac{2}{3},&
\mathrm{H}^{(0)-1}_{3 1} &= \frac{3i}{5},& 
\mathrm{H}^{(0)-1}_{3 3} &= \frac{2i}{5}, \nonumber \\
\mathrm{H}^{(0)-1}_{4 0} &= \frac{1}{5},& 
\mathrm{H}^{(0)-1}_{4 2} &= \frac{4}{7},& 
\mathrm{H}^{(0)-1}_{4 4} &= \frac{8}{35}, \nonumber \\
\mathrm{H}^{(1)-1}_{1 1} &= -1,& 
\mathrm{H}^{(1)-1}_{2 2} &= \frac{i}{3},& 
\mathrm{H}^{(1)-1}_{3 1} &= \frac{6}{5}, \nonumber \\ 
\mathrm{H}^{(1)-1}_{3 3} &= \frac{2}{15}, &
\mathrm{H}^{(1)-1}_{4 2} &= -\frac{10i}{21},& 
\mathrm{H}^{(1)-1}_{4 4} &= -\frac{2i}{35}, \nonumber \\
\mathrm{H}^{(2)-1}_{2 2} &= \frac{1}{3},& 
\mathrm{H}^{(2)-1}_{3 3} &= -\frac{i}{15},& 
\mathrm{H}^{(2)-1}_{4 2} &= -\frac{5}{7}, \nonumber \\ 
\mathrm{H}^{(2)-1}_{4 4} &= -\frac{2}{105}, && && \nonumber \\
\mathrm{H}^{(3)-1}_{3 3} &= -\frac{1}{15},& 
\mathrm{H}^{(3)-1}_{4 4} &= \frac{i}{105}, && \nonumber \\
\mathrm{H}^{(4)-1}_{4 4} &= \frac{1}{105}. && &&
\end{align}

\bibliography{tidal_forcing}{}
\bibliographystyle{mn2e}

\end{document}